\pdfoutput=1
\RequirePackage{ifpdf}
\ifpdf 
\documentclass[pdftex]{sigma}
\else
\documentclass{sigma}
\fi

\usepackage{bm}
\usepackage{mathtools}

\numberwithin{equation}{section}

\newcommand {\ms}{{\mathstrut}}

\newcommand {\bbC}{\mathbb C}

\newcommand {\bbL}{\mathbb L}
\newcommand {\tbbL}{\widetilde{\mathbb L \hskip -.25em} \hskip .25em}
\newcommand {\bbM}{\mathbb M}
\newcommand {\tbbM}{\widetilde{\mathbb M}}
\newcommand {\bbN}{\mathbb N}

\newcommand {\bbZ}{\mathbb Z}

\newcommand {\calB}{\mathcal B}

\newcommand {\calG}{\mathcal G}
\newcommand {\calL}{\mathcal L}
\newcommand {\calM}{\mathcal M}
\newcommand {\calO}{\mathcal O}
\newcommand {\calQ}{\mathcal Q}
\newcommand {\calR}{\mathcal R}
\newcommand {\calT}{\mathcal T}

\newcommand {\gothg}{\mathfrak g}
\newcommand {\gothgl}{\mathfrak{gl}}
\newcommand {\gothh}{\mathfrak h}
\newcommand {\gothk}{\mathfrak k}
\newcommand {\gothsl}{\mathfrak{sl}}
\newcommand {\hgothh}{\widehat{\mathfrak h}}
\newcommand {\tgothh}{\widetilde{\mathfrak h}}

\newcommand {\hi}{\widehat I}

\newcommand {\sllpo}{\mathfrak{sl}_{l+1}}
\newcommand {\lsllpo}{{\mathcal L}({\mathfrak{sl}}_{l+1})}
\newcommand {\tlsllpo}{\widetilde{\mathcal L}({\mathfrak{sl}}_{l+1})}
\newcommand {\hlsllpo}{\widehat{\mathcal L}(\mathfrak{sl}_{l + 1})}

\newcommand {\uqlsllpo}{\mathrm U_q(\mathcal L(\mathfrak{sl}_{l + 1}))}
\newcommand {\uqtlsllpo}{\mathrm U_q(\widetilde{\mathcal L}(\mathfrak{sl}_{l + 1}))}
\newcommand {\uqhlsllpo}{\mathrm U_q(\widehat{\mathcal L}(\mathfrak{sl}_{l + 1}))}

\newcommand {\uqbm}{\mathrm U_q(\mathfrak b_-)}
\newcommand {\uqbp}{\mathrm U_q(\mathfrak b_+)}

\newcommand {\uqgllpo}{\mathrm U_q(\mathfrak{gl}_{l + 1})}

\newcommand {\uqg}{\mathrm U_q(\mathfrak g)}

\newcommand {\uqwtLsllpo}{\mathrm U_q(\widetilde{\mathcal L}(\mathfrak{sl}_{l+1}))}

\newcommand {\uqsllpo}{\mathrm U_q(\mathfrak{sl}_{l + 1})}

\newcommand {\mbar}[3]{\hskip #2 \overline{\hskip -#2 #1 \hskip -#3} \hskip #3}

\newcommand {\ovcalQ}{\mbar{\mathcal Q}{.1em}{.03em}}
\newcommand {\ovlambda}{\mbar{\lambda}{.03em}{.03em}}
\newcommand {\ovotimes}{\mathbin{\mbar{\otimes}{.1em}{.1em}}}
\newcommand {\ovPsi}{\mbar{\Psi}{.03em}{.03em}}
\newcommand {\ovrho}{\mbar{\rho}{.03em}{.03em}}
\newcommand {\ovtheta}{\mbar{\theta}{.03em}{.03em}}
\newcommand {\ovv}{\mbar{v}{.05em}{.01em}}
\newcommand {\ovxi}{\mbar{\xi}{.05em}{.03em}}

\newcommand {\rme}{\mathrm e}

\DeclareMathOperator {\End}{End}
\DeclareMathOperator {\id}{\mathrm{id}}
\DeclareMathOperator {\Mat}{Mat}
\DeclareMathOperator {\Osc}{Osc}
\DeclareMathOperator {\tr}{\mathrm{tr}}

\begin{document}

\allowdisplaybreaks

\newcommand{\arXivNumber}{1702.08710}

\renewcommand{\thefootnote}{}

\renewcommand{\PaperNumber}{043}

\FirstPageHeading

\ShortArticleName{Highest $\ell$-Weight Representations and Functional Relations}

\ArticleName{Highest $\boldsymbol{\ell}$-Weight Representations\\ and Functional Relations\footnote{This paper is a~contribution to the Special Issue on Recent Advances in Quantum Integrable Systems. The full collection is available at \href{http://www.emis.de/journals/SIGMA/RAQIS2016.html}{http://www.emis.de/journals/SIGMA/RAQIS2016.html}}}

\Author{Khazret S.~NIROV~$^{\dag\ddag}$ and Alexander V.~RAZUMOV~$^\S$}

\AuthorNameForHeading{Kh.S.~Nirov and A.V.~Razumov}

\Address{$^\dag$~Institute for Nuclear Research of the Russian Academy of Sciences,\\
\hphantom{$^\dag$}~60th October Ave.~7a, 117312 Moscow, Russia}

\Address{$^\ddag$~Mathematics and Natural Sciences, University of Wuppertal, 42097 Wuppertal, Germany}
\EmailD{\href{mailto:nirov@uni-wuppertal.de}{nirov@uni-wuppertal.de}}

\Address{$^\S$~Institute for High Energy Physics, NRC ``Kurchatov Institute'',\\
\hphantom{$^\S$}~142281 Protvino, Mos\-cow region, Russia}
\EmailD{\href{mailto:Alexander.Razumov@ihep.ru}{Alexander.Razumov@ihep.ru}}

\ArticleDates{Received March 01, 2017, in f\/inal form June 06, 2017; Published online June 17, 2017}

\Abstract{We discuss highest $\ell$-weight representations of quantum loop algebras and the corresponding functional relations between integrability objects. In particular, we compare the prefundamental and $q$-oscillator representations of the positive Borel subalgebras of the quantum group $\mathrm{U}_q(\mathcal L(\mathfrak{sl}_{l+1}))$ for arbitrary values of $l$. Our article has partially the nature of a~short review, but it also contains new results. These are the expressions for the $L$-operators, and the exact relationship between dif\/ferent representations, as a byproduct resulting in certain conclusions about functional relations.}


\Classification{17B37; 16T25; 17B10}

\renewcommand{\thefootnote}{\arabic{footnote}}
\setcounter{footnote}{0}

\section{Introduction}

The use of highest $\ell$-weights and highest $\ell$-weight vectors allows one to properly ref\/ine the spectral data about highest weight representations in the same way as the generalized eigenvalues and eigenvectors do for the eigenvalue problems. The corresponding notion proved especially useful in the classif\/ication of irreducible f\/inite-dimensio\-nal~\cite{ChaPre91, ChaPre94} and inf\/inite-dimen\-sio\-nal~\mbox{\cite{HerJim12, MukYou14}} representations of quantum loop algebras and their Borel subalgebras. For quantum af\/f\/ine algebras and their Borel subalgebras, the related category of representations was studied in~\cite{Her07} and~\cite{HerJim12}, respectively.

The study of dif\/ferent representations of quantum groups in application to quantum integrable systems received new impetus from the remarkable papers by Bazhanov, Lukyanov and Zamolodchikov \cite{BazLukZam96, BazLukZam97, BazLukZam99}. In general terms, in the approach advanced in these papers, the investigation of quantum integrable systems is reduced to the study of representations of the corresponding quantum groups. More specif\/ically, the method is based on the universal $R$-matrix. By def\/inition, it is an element of the tensor product of two copies of the quantum group, and one calls the representation spaces for the f\/irst and second factors of this tensor product the auxiliary and quantum spaces, respectively. Here, a representation of the quantum group in the auxiliary space gives an integrability object which is either a monodromy- or a transfer-type operator. A representation in the quantum space def\/ines then a physical model. For example, it can be a low-dimensional quantum f\/ield theory as in~\cite{BazHibKho02, BazLukZam96, BazLukZam97, BazLukZam99} or a spin-chain model as in~\cite{ BooGoeKluNirRaz14b, BooGoeKluNirRaz14a,NirRaz16a}. The integrability objects satisfy functional relations as a consequence of the cha\-rac\-teristics of the representations of the quantum group in the auxiliary and quantum spaces. In fact, such functional relations can be derived in a {\em universal form\/}, f\/ixing representations of the quantum group only in the auxiliary space and being thus independent of the representations in the quantum space. We would like to refer to the paper~\cite{BooGoeKluNirRaz14a} for more details.

Now, it is relevant to point out that the universal $R$-matrix is actually an element of (a~completion of) the tensor product of the positive and negative Borel subalgebras of the initial quantum group. This means inter alia that given a representation of the whole quantum group, one can obtain representations for the construction of integrability objects by restricting it to the corresponding Borel subalgebras. This way does work for the monodromy and transfer ope\-ra\-tors. However, there are representations of the Borel subalgebras which cannot be obtained by such a simple restriction. In particular, one obtains such representations mapping the positive Borel subalgebra to a $q$-oscillator algebra and using representations of the latter. This method was proposed in \cite{BazLukZam97, BazLukZam99} for the construction of CFT analogs of the Baxter's $Q$-operators. Such representations can be deduced from those ones employed earlier for the construction of monodromy and transfer operators by a certain degeneration procedure \cite{BazHibKho02, BooGoeKluNirRaz13,BooGoeKluNirRaz14b, BooGoeKluNirRaz14a, NirRaz16a}. This relationship between representations ascertains that the corresponding integrability objects are involved in nontrivial functional equations.

In \cite{HerJim12}, Hernandez and Jimbo studied inductive limits of the Kirillov--Reshetikhin modules and obtained new simple inf\/inite-dimensional representations of the Borel subalgebras of quantum loop algebras. These are highest $\ell$-weight modules characterized by highest $\ell$-weights of simplest possible form. Later on, in \cite{FreHer15}, based on the notion of $q$-characters, generalized Baxter's $TQ$-relations were given an interpretation as of relations in the Grothendieck ring of the category $\calO$ from \cite{HerJim12}. Just as in \cite{FreHer15}, we call the above highest $\ell$-weight representations {\em prefundamental\/}.

In the paper \cite{BooGoeKluNirRaz16}, we found the $\ell$-weights and the corresponding $\ell$-weight vectors for the f\/inite- and inf\/inite-dimensional representations of the quantum loop algebra $\uqlsllpo$ for $l = 1$ and~$2$ constructed through Jimbo's evaluation representations. We also found there the $\ell$-weights and the $\ell$-weight vectors for the $q$-oscillator representations of the positive Borel subalgebras of the same quantum groups. The work \cite{BooGoeKluNirRaz16} showed how the $q$-oscillator and prefundamental representations are explicitly related. However, the consideration of these cases, with $l = 1$ and $l = 2$, did not allow for a direct generalization to the arbitrary higher ranks. Quite recently, based on the paper \cite{NirRaz16b}, we have considered the general case with arbitrary $l$ and obtained the $\ell$-weights and the corresponding $\ell$-weight vectors for $q$-oscillator representations of the positive Borel subalgebra of $\uqlsllpo$~\cite{BooGoeKluNirRaz17}. Here we use the notations and calculations of \cite{BooGoeKluNirRaz17}.

The article has partially the nature of a short review, but it also contains new results (see Sections~\ref{s:aafr} and~\ref{s:hlwfr}). In Section~\ref{s:qg}, we recall the quantum groups in general in order to introduce the universal $R$-matrix as the main tool. Applying to it dif\/ferent representations, we def\/ine the universal integrability objects and discuss their basic properties. In Section~\ref{s:qglrs}, we specify the general notion of quantum groups to the quantum group of the general linear Lie algebra of arbitrary rank. Here we discuss its highest weight representations. In Section~\ref{s:qlars}, we describe the quantum group of the untwisted loop algebra of the special linear Lie algebra of arbitrary rank. It is traditional to call this object a quantum loop algebra. We construct representations of this algebra using the corresponding Jimbo's evaluation homomorphism. Besides, we def\/ine the Borel subalgebras of the quantum loop algebra and, following our paper~\cite{NirRaz16b}, describe their representations. In Section~\ref{s:hlwr}, we recall necessary data on the highest $\ell$-weight representations with rational $\ell$-weights. As the respective basic examples of our special interest, we discuss the prefundamental and $q$-oscillator representations of the positive Borel subalgebra of the quantum loop algebra under consideration. In Section~\ref{s:aafr}, we describe the symmetry transformations which allow us to construct more $q$-oscillator representations of the Borel subalgebra. In Section~\ref{s:hlwfr}, we present the highest $\ell$-weights for the $q$-oscillator representations from the preceding section and discuss explicit relations between them. These relations reproduce the def\/ining characteristics of the functional relations between the universal integrability objects. Our results establish a~direct connection between the $q$-oscillator and prefundamental representations. We conclude with some remarks.

To subsequently def\/ine the quantum groups, we introduce the corresponding deformation parameter. Here we determine a nonzero complex number~$\hbar$, such that $q = \exp \hbar$ is not a root of unity. With such a deformation parameter~$q$, the quantum groups under consideration are treated as $\bbC$-algebras. Besides, we assume that
\begin{gather*}
q^\nu = \exp (\hbar \nu), \qquad \nu \in \bbC.
\end{gather*}
For the $q$-numbers and $q$-factorials we use the traditional notations
\begin{gather*}
[n]_q = \frac{q^n - q^{-n}}{q - q^{-1}}, \qquad n \in \bbZ,
\end{gather*}
and
\begin{gather*}
[n]_q! = \prod_{m = 1}^n [m]_q, \quad n \in \bbN, \qquad [0]_q! = 1,
\end{gather*}
respectively. We also use the convenience of the notation
\begin{gather*}
\kappa_q = q - q^{-1}.
\end{gather*}

\section{Quantum groups and integrability objects} \label{s:qg}

Following Drinfeld \cite{Dri85, Dri87} and Jimbo \cite{Jim85}, we treat a quantum group $\calG$ as a one-parameter deformation of the universal enveloping algebra of a Lie algebra $\gothg$. Hence the usual notation for $\calG$ as $\uqg$, where $q$ is the mentioned deformation parameter. The nature of the quantum group can essentially depend
on the specif\/ication of this parameter, see the books \cite{ChaPre94, JimMiw95, EtiFreKir98} for a discussion of the point. The quantum group is def\/ined as a Hopf algebra with respect to appropriate co-multiplication $\Delta$, antipode $S$ and co-unit $\epsilon$. It is also a Hopf algebra with respect to the opposite co-multiplication $\Delta^{\mathrm{op}} = \Pi \circ \Delta$, where the permutation operator is def\/ined by
\begin{gather*}
\Pi (a \otimes b) = b \otimes a, \qquad a, b \in \calG.
\end{gather*}
The quantum group is a quasitriangular Hopf algebra. It means that there exists the so-called universal R-matrix $\calR$ being an element of the completed tensor product of two copies of the quantum group and relating the co-multiplication and the opposite one as
\begin{gather*}
\Delta^{\mathrm{op}}(a) = \calR \Delta(a) \calR^{-1}, \qquad a \in \calG, \qquad \calR \in \calG \otimes \calG,
\end{gather*}
and satisfying the following relations:
\begin{gather*}
(\Delta \otimes \id) (\calR) = \calR^{13} \calR^{23}, \qquad (\id \otimes \Delta) (\calR) = \calR^{13} \calR^{12},
\end{gather*}
where the indices have the standard meaning. The above relations lead to the following equation for the universal $R$-matrix:
\begin{gather*}
\calR^{12} \calR^{13} \calR^{23} = \calR^{23} \calR^{13} \calR^{12}
\end{gather*}
called the Yang--Baxter equation. It is def\/ined in the tensor cube of the quantum group. However, it is important to note that the universal $R$-matrix belongs to the completed tensor product of the positive and the negative Borel subalgebras
of the quantum group,
\begin{gather*}
\calR \in \calB_+ \otimes \calB_- \subset \calG \otimes \calG.
\end{gather*}
This fact has profound implications in the theory of quantum integrable systems. First of all, it means that the Yang--Baxter equation is actually def\/ined not in the full tensor cube of the quantum group, but in the space $\calB_+ \otimes \calG \otimes \calB_-$. Secondly, it allows one to consider integrability objects, such as monodromy- and transfer-type operators, having essentially dif\/ferent nature.

To be more specif\/ic, let us describe how the integrability objects associated with the quantum group $\calG$ and its Borel subalgebra $\calB_+$ arise in general. We refer the reader to \cite{BooGoeKluNirRaz13} for more details. With the help of a group-like element $t$, by def\/inition satisfying the relation
\begin{gather*}
\Delta(t) = t \otimes t,
\end{gather*}
we obtain from the Yang--Baxter equation the equation
\begin{gather}
\big(\calR^{13} t^1\big) \big(\calR^{23} t^2\big) = \big(\calR^{12}\big)^{-1} \big(\calR^{23} t^2\big) \big(\calR^{13} t^1\big) \big(\calR^{12}\big). \label{1.1}
\end{gather}
Let $\varphi$ be a representation of $\calG$ in a vector space $V$. We def\/ine the monodromy-type operator $\calM_\varphi$ associated with this representation as
\begin{gather*}
\calM_\varphi = (\varphi \otimes \id) (\calR)
\end{gather*}
and see that it is an element of $\mathrm{End}(V) \otimes \calB_-$. Next we def\/ine the corresponding transfer-type ope\-ra\-tor~$\calT_\varphi$ as
\begin{gather*}
\calT_\varphi = (\tr_V \otimes \id) (\calM_\varphi (\varphi(t) \otimes 1)) =
((\tr_V \circ \varphi) \otimes \id)(\calR(t \otimes 1)).
\end{gather*}
Here $1$ is the unit element of $\calG$, and we assume that $t$ is such that the trace over the representation space $V$ is well-def\/ined. It is clear that $\calT_\varphi$ belongs to the negative Borel subalgebra $\calB_- \subset \calG$. We see that to def\/ine these integrability objects, $\calM_\varphi$ and $\calT_\varphi$, one starts with a representation of the whole quantum group $\calG$, but one then uses only its restriction to the positive Borel subalgebra~$\calB_+$. Moreover, one can def\/ine in this way dif\/ferent transfer-type operators associated with representations~$\varphi_1$ and~$\varphi_2$ of $\calG$ and see directly from (\ref{1.1}) that they commute,
\begin{gather*}
\calT_{\varphi_1} \calT_{\varphi_2} = \calT_{\varphi_2} \calT_{\varphi_1}.
\end{gather*}
This is the primary indication of the integrability of models which can be associated with $\calG$. One can also consider parameterized representations of the quantum group and arrive at the commutativity of the transfer-type
operators for dif\/ferent values of the corresponding parameters.

To have more integrability objects, one considers representations which are not restrictions of a representation of $\calG$ to $\calB_+$ or $\calB_-$, and which cannot be extended from the Borel subalgebras to a representation
of the full quantum group. Let $\rho$ be such a representation of $\calB_+$ in a vector space $W$. We introduce a monodromy-type operator $\calL_\rho$ associated with this representation,
\begin{gather*}
\calL_\rho = (\rho \otimes \id) (\calR),
\end{gather*}
being an element of $\mathrm{End}(W) \otimes \calB_-$. The corresponding $Q$-operator $\calQ_\rho$ is then an element of $\calB_-$ def\/ined as
\begin{gather*}
\calQ_\rho = (\tr_W \otimes \id) (\calL_\rho (\rho(t) \otimes 1)) =
((\tr_W \circ \rho) \otimes \id)(\calR(t \otimes 1)).
\end{gather*}
Here again, $1$ is the unity of $\calG$, and the group-like element $t$ is such that the trace over the representation space $W$ is well-def\/ined. Using equation (\ref{1.1}), one can show that
\begin{gather*}
\calQ_{\rho} \calT_{\varphi} = \calT_{\varphi} \calQ_{\rho}.
\end{gather*}
Appropriate representations of such kind to be used for $\rho$, the so-called $q$-oscillator representations, were considered for the f\/irst time in \cite{BazLukZam97, BazLukZam99} and \cite{BazHibKho02}, where integrable structures of conformal quantum f\/ield theories were investigated. One can show that also the $Q$-operators commute for dif\/ferent representations $\rho_1$ and $\rho_2$. However, this commutativity and other nontrivial relations between the transfer-type integrability objects do not follow simply from the Yang--Baxter equation~(\ref{1.1}) anymore. For some partial cases the commutativity was proved exploring details on the tensor products of the respective representations in the papers \cite{BazHibKho02, BooGoeKluNirRaz13, BooGoeKluNirRaz14b,BooGoeKluNirRaz14a, NirRaz16a}. The proof for the general case was given in the paper \cite[Section~5.2]{FreHer15}.

It is convenient to use a more general def\/inition of monodromy-type operators. Here the mappings $\varphi$ and $\rho$ are homomorphisms from $\calG$, or $\calB_+$, to some algebra with a relevant set of representations. One constructs integrability objects with such $\varphi$ and $\rho$ and then apply to them appropriate representations.

We note f\/inally that the integrability objects above have been introduced in such a way that only a representation of the quantum group in the auxiliary space was f\/ixed, and no representation in the quantum space was chosen. In this sense, they are model independent. For this reason we call the above monodromy- and transfer-type operators the universal integrability objects. To obtain the corresponding integrability objects for specif\/ic models, one has to f\/ix a representation of the quantum group in the quantum space.

Further, we need to specify the quantum group $\calG = \uqg$ and its Borel subalgebras. Actually, we will consider two cases with the Lie algebra~$\gothg$ being the general linear Lie algebra $\gothgl_{l+1}$ and the loop algebra~$\lsllpo$. Recall that the deformation parameter $q$ is always supposed to be an exponential of a complex number~$\hbar$, such that~$q$ is not a root of unity. It allows one to treat~$\uqg$ as a unital associative
$\bbC$-algebra obtained by the $q$-deformation of the universal enveloping algebra of the Lie algebra~$\gothg$.

\section[Quantum group $\uqgllpo$ and its representations]{Quantum group $\boldsymbol{\uqgllpo}$ and its representations}\label{s:qglrs}

Let $\gothk_{l+1}$ be the standard Cartan subalgebra of the Lie algebra $\gothgl_{l+1}$ and $\triangle$ be the root system of $\gothgl_{l+1}$ respective to $\gothk_{l+1}$. Denote by $\{\alpha_i \in \gothk_{l+1}^* \,|\, i = 1,\ldots,l\}$ the corresponding set of simple roots. Then, with $K_i$, $i = 1,\ldots,l+1$, forming the standard basis of $\gothk_{l+1}$, we have
\begin{gather*}
\langle \alpha_j , K_i \rangle = c_{i j},
\end{gather*}
where
\begin{gather*}
c_{i j} = \delta_{i j} - \delta_{i, j + 1}.
\label{e:2.1}
\end{gather*}
The general linear Lie algebra $\gothgl_{l+1}$ is generated by $2 l$ Chevalley generators $E_i$, $F_i$, $i = 1, \ldots, l$, and by $l + 1$ Cartan elements $K_i$, together satisfying well-known def\/ining relations supplemented also with the Serre relations. For the total root system we have
\begin{gather*}
\triangle = \triangle_+ \sqcup \triangle_-,
\end{gather*}
where $\triangle_+$ is the set of positive roots which are all of the form
\begin{gather*}
\alpha_{i j} = \sum_{k = i}^{j-1} \alpha_k, \qquad 1 \le i < j \le l + 1,
\end{gather*}
and we also have $\alpha_i = \alpha_{i, i+1}$, and $\triangle_- = - \triangle_+$ is the set of negative roots. The restriction to the special linear Lie algebra $\gothsl_{l+1}$ is obtained by setting
\begin{gather*}
H_i = K_i - K_{i + 1}, \qquad i = 1,\ldots,l,
\end{gather*}
as the generators of the standard Cartan subalgebra $\gothh_{l+1}$ of $\gothsl_{l+1}$ and keeping $E_i$ and $F_i$ as the corresponding Chevalley generators. The positive and negative roots of $\gothsl_{l+1}$ are the restrictions of $\alpha_{ij}$ and $-\alpha_{ij}$ to $\gothh_{l+1}$, respectively. Then we have
\begin{gather*}
\langle \alpha_j , H_i \rangle = a_{i j},
\end{gather*}
where
\begin{gather*}
a_{ij} = c_{ij} - c_{i+1, j}
\end{gather*}
are the entries of the Cartan matrix of $\gothsl_{l+1}$. As usual the fundamental weights $\omega_i \in \gothh^*$, $i = 1, \ldots, l$, are def\/ined by the relations
\begin{gather*}
\langle \omega_i, H_j \rangle = \delta_{i j}, \qquad j = 1, \ldots, l.
\end{gather*}

The quantum group $\uqgllpo$ is generated by the elements
\begin{gather*}
E_i, \quad F_i, \quad i = 1,\ldots,l, \qquad q^{X}, \quad X \in \gothk_{l+1},
\end{gather*}
satisfying the def\/ining relations
\begin{gather}
q^0 = 1, \qquad q^{X_1} q^{X_2} = q^{X_1 + X_2}, \label{2.8a} \\
q^X E_i q^{-X} = q^{\langle \alpha_i, X \rangle} E_i, \qquad q^X F_i q^{-X} = q^{-\langle \alpha_i, X \rangle} F_i, \label{2.11a} \\
[E_i, F_j] = \delta_{ij} \frac{q^{K_i - K_{i+1}} - q^{- K_i + K_{i + 1}}}{q - q^{-1}},
\label{2.10a}
\end{gather}
and the Serre relations
\begin{gather*}
E_i E_j = E_j E_i, \qquad F_i F_j = F_j F_i, \qquad |i - j| \ge 2, \\
E_i^2 E_{i \pm 1} - [2]_q E_i E_{i \pm 1} E_i + E_{i \pm 1} E_i^2 = 0, \qquad F_i^2 F_{i \pm 1} - [2]_q F_i F_{i \pm 1} F_i + F_{i \pm 1} F_i^2 = 0.
\end{gather*}
The set of the elements of the form $q^X$ is parameterized by the Cartan subalgebra~$\gothk_{l+1}$. The quantum group $\uqsllpo$ is generated by the same generators as~$\uqgllpo$, only that the generators $q^X$ of $\uqsllpo$ are parameterized by the Cartan subalgebra $\gothh_{l+1}$. The generators of $\uqsllpo$ fulf\/il the same relations as of~$\uqgllpo$, with one exception that (\ref{2.10a}) takes now the form
\begin{gather*}
[E_i, F_j] = \delta_{ij} \frac{q^{H_i} - q^{-H_i}}{q - q^{-1}}.
\end{gather*}
We assume everywhere that
\begin{gather*}
q^{X + \nu} = q^\nu q^X, \qquad [ X + \nu ]_q = \frac{q^{X+\nu} - q^{-X-\nu}}{q - q^{-1}}, \qquad X \in \gothk_{l+1}, \quad \nu \in \bbC.
\end{gather*}

Both quantum groups, $\uqgllpo$ and $\uqsllpo$, are Hopf algebras with respect to the co-multiplication, antipode and co-unit def\/ined as follows:
\begin{gather*}
\Delta(q^X) = q^X \otimes q^X, \qquad \Delta(E_i) = E_i \otimes 1 + q^{H_i} \otimes E_i, \qquad \Delta(F_i) = F_i \otimes q^{-H_i} + 1 \otimes F_i, \\
S(q^X) = q^{- X}, \qquad S(E_i) = - q^{- H_i} E_i, \qquad S(F_i) = - F_i q^{H_i},\\
\epsilon(q^X) = 1, \qquad \epsilon(E_i) = 0, \qquad \epsilon(F_i) = 0.
\end{gather*}
Although these relations are not used in this paper, we note that the Hopf algebra structure is crucial for the quantum integrable systems associated with these quantum groups.

The quantum group $\uqgllpo$ possesses a Poincar\'e--Birkhof\/f--Witt basis. To construct it, one needs an appropriate def\/inition of the root vectors. We f\/irst introduce a $Q$-gradation of $\uqgllpo$ with respect to the root lattice of $\gothgl_{l+1}$. The latter is the abelian group
\begin{gather*}
Q = \bigoplus_{i=1}^{l} \bbZ \alpha_i,
\end{gather*}
and $\uqgllpo$ becomes $Q$-graded if we assume that
\begin{gather*}
E_i \in \uqgllpo_{\alpha_i}, \qquad F_i \in \uqgllpo_{-\alpha_i}, \qquad q^X \in \uqgllpo_{0}
\end{gather*}
for all $i = 1,\ldots,l$ and $X \in \gothk_{l+1}$. An element $a$ of $\uqgllpo$ is called a root vector corresponding to the root~$\gamma$ of~$\gothgl_{l+1}$ if $a \in \uqgllpo_\gamma$. In the case under consideration, this is equivalent to
the relations
\begin{gather*}
q^X a q^{-X} = q^{\langle \gamma , X \rangle} a, \qquad X \in \gothk_{l+1}.
\end{gather*}
The Chevalley generators $E_i$ and $F_i$ are obviously root vectors corresponding to the roots~$\alpha_i$ and~$-\alpha_i$. Now we def\/ine the whole set of linearly independent root vectors. We start introducing the set
\begin{gather*}
\Lambda_l = \{(i, j) \in \bbN \times \bbN \,|\, 1 \le i < j \le l + 1 \}
\end{gather*}
and def\/ine the elements $E_{ij}$ and $F_{ij}$, $(i, j) \in \Lambda_l$, according to Jimbo \cite{Jim86a},
\begin{gather*}
E_{i,i+1} = E_i, \quad i = 1,\ldots,l, \qquad E_{i j} = E_{i, j - 1} E_{j - 1, j} - q E_{j - 1, j} E_{i, j - 1}, \quad j - i > 1, \\
F_{i, i + 1} = F_i, \quad i = 1,\ldots,l, \qquad F_{i j} = F_{j - 1, j} F_{i, j - 1} - q^{-1} F_{i, j - 1} F_{j - 1, j}, \quad j - i > 1.
\end{gather*}
The elements $E_{ij}$ are the root vectors corresponding to the positive roots~$\alpha_{ij}$, and the elements~$F_{ij}$ are the root vectors corresponding to the negative roots~$-\alpha_{ij}$. The Cartan--Weyl generators of~$\uqgllpo$ are the elements $q^X$, $X \in \gothk_{l+1}$, and $E_{ij}$, $F_{ij}$. The Poincar\'e--Birkhof\/f--Witt basis of~$\uqgllpo$ is formed by the ordered monomials constructed from the Cartan--Weyl generators. To def\/ine such monomials explicitly, let us impose the lexicographic order on the set $\Lambda_l$, which means that $(i, j) < (m, n)$ if $i < m$, or if $i = m$ and $j < n$.\footnote{Note that in~\cite{NirRaz16b} we used the co-lexicographic ordering. If we def\/ine an ordering of the positive roots according to the co-lexicographic order on $\Lambda_l$, we would have a normal ordering in the sense of \cite{AshSmiTol79, LezSav74}. Here, in contrast, we use the lexicographic order and obtain a dif\/ferent realization of the normal ordering of positive roots.} Then, a respectively ordered monomial being an appropriate Poincar\'e--Birkhof\/f--Witt basis element can be taken in the form
\begin{gather}
F_{i_1 j_1} \cdots F_{i_r j_r} q^X E_{m_1 n_1} \cdots E_{m_s n_s}, \label{fqxe}
\end{gather}
where $(i_1, j_1) \le \cdots \le (i_r, j_r)$, $(m_1, n_1) \le \cdots \le (m_s, n_s)$ and $X$ is an arbitrary element of~$\gothk_{l+1}$. The monomials of the same form with $X \in \gothh_{l+1}$ form a Poincar\'e--Birkhof\/f--Witt basis of~$\uqsllpo$.

By def\/inition of the Poincar\'e--Birkhof\/f--Witt basis, any monomial can be given by a sum of ordered monomials of the form (\ref{fqxe}). To f\/ind an ordered form of a given monomial, one must be able to reorder, if necessary, the constituent elements $q^X$, $E_{ij}$ and $F_{ij}$, and the process of reordering requires certain relations between these elements. All such relations were derived in~\cite{Yam89}. In a recent paper~\cite{NirRaz16b}, we adopted those relations in a suitable for our def\/initions form and used them to obtain the def\/ining relations of the Verma $\uqgllpo$-module.

We denote by $\widetilde V^\lambda$ the Verma $\uqgllpo$-module corresponding to the highest weight $\lambda \in \gothk_{l+1}^*$. Here, for the highest weight vector $v^\lambda$ we have the def\/ining relations
\begin{gather*}
E_i v^\lambda = 0, \quad i = 1,\ldots,l, \qquad q^X v^\lambda = q^{\langle \lambda, X \rangle} v^\lambda, \quad X \in \gothk_{l+1}, \quad \lambda \in \gothk_{l+1}^*. \label{3.1}
\end{gather*}
As usual, the highest weight is identif\/ied with its components respective to the basis of $\gothk_{l+1}$,
\begin{gather*}
\lambda_i = \langle \lambda, K_i \rangle.
\end{gather*}
The representation of $\uqgllpo$ corresponding to $\widetilde V^\lambda$ is denoted by~$\widetilde \pi^\lambda$. The structure and properties of $\widetilde V^\lambda$ and~$\widetilde \pi^\lambda$ for $l = 1$ and $l = 2$ are considered in much detail in our papers \cite{BooGoeKluNirRaz13, BooGoeKluNirRaz14b,BooGoeKluNirRaz14a, BooGoeKluNirRaz16, NirRaz16a}. The case of general~$l$ was studied in our recent paper~\cite{NirRaz16b}. Here we shortly recall the corresponding results from~\cite{NirRaz16b}.

Let us denote by $\bm{m}$ the $l(l+1)/2$-tuple of non-negative integers $m_{ij}$, arranged in the lexicographic order of $(i, j) \in \Lambda_l$. Explicitly we have
\begin{gather*}
{\bm m} = (m_{12}, m_{13}, \ldots, m_{1,l+1}, \ldots, m_{i, i+1}, m_{i, i+2}, \ldots, m_{i, l + 1}, \ldots, m_{l, l + 1}).
\end{gather*}
The vectors
\begin{gather*}
v_{\bm m} = F_{12}^{m_{12}} F_{13}^{m_{13}} \cdots F_{1, l + 1}^{m_{1, l + 1}} \cdots F_{i, i + 1}^{m_{i, i + 1}} \cdots F_{i, i + 2}^{m_{i, i + 2}} \cdots
 F_{i, l + 1}^{m_{i, l + 1}} \cdots F_{l, l + 1}^{m_{l, l + 1}} v_{\bm 0},
\end{gather*}
where for consistency $v_{\bm0}$ denotes the highest-weight vector $v^\lambda$, form a basis of $\widetilde V^\lambda$. Note that in~\cite{NirRaz16b} the integers~$m_{ij}$ were arranged in the co-lexicographic order, but the basis vectors~$v_{\bm{m}}$ for both orderings, the lexicographic and co-lexicographic ones, coincide, and the def\/ining module relations do not distinguish the choice between these orderings.

The $\uqgllpo$-module def\/ining relations from \cite{NirRaz16b} are as follows:
\begin{gather}
 q^{\nu K_i} v_{\bm m} = q^{\nu \big(\lambda_i + \sum\limits_{k=1}^{i-1} m_{ki}
- \sum\limits_{k=i+1}^{l+1} m_{ik}\big)} v_{\bm m}, \qquad i = 1,\ldots,l+1,\label{3.3} \\
 F_{i, i + 1} v_{\bm m} = q^{- \sum\limits_{k = 1}^{i-1} (m_{k i} - m_{k, i + 1})} v_{{\bm m} + \epsilon_{i, i + 1}} + \sum_{j = 1}^{i - 1} q^{- \sum\limits_{k=1}^{j - 1}(m_{k i} - m_{k, i+1})} [m_{j i}]_q v_{{\bm m} - \epsilon_{j i} + \epsilon_{j, i + 1}}, \notag \\
 E_{i, i + 1} v_{\bm m} = \bigg[\lambda_i - \lambda_{i+1} - \sum_{j = i + 2}^{l + 1}(m_{i j} - m_{i + 1, j})- m_{i, i + 1} + 1\bigg]_q [m_{i, i + 1}]_q v_{{\bm m} - \epsilon_{i, i + 1}}\notag \\
 \hphantom{E_{i, i + 1} v_{\bm m} =}{} + q^{\lambda_i - \lambda_{i + 1} - 2 m_{i, i + 1}
- \sum\limits_{j = i + 2}^{l + 1} (m_{i j} - m_{i + 1, j})}
\sum\limits_{j = 1}^{i - 1} q^{\sum\limits_{k = j + 1}^{i - 1} (m_{k i} - m_{k, i + 1})}
[m_{j, i + 1}]_q v_{{\bm m} - \epsilon_{j, i + 1} + \epsilon_{j i}} \notag \\
\hphantom{E_{i, i + 1} v_{\bm m} =}{} - \sum_{j = i + 2}^{l + 1} q^{- \lambda_i + \lambda_{i + 1} - 2 + \sum\limits_{k = j}^{l + 1} (m_{i k} - m_{i + 1, k})} [m_{i j}]_q v_{{\bm m} - \epsilon_{i j} + \epsilon_{i + 1, j}},\notag
\end{gather}
where $i = 1, \ldots, l$ in last two equations. Here and below ${\bm m} + \nu \epsilon_{i j}$ means shifting by $\nu$ the entry~$m_{i j}$ in the $l(l + 1)/2$-tuple ${\bm m}$. In what follows, we will also need the action of the root vec\-tors~$F_{1, l+1}$ on the basis vec\-tors~$v_{\bm m}$. This is given by the equation
\begin{gather*}
F_{1, l + 1} v_{\bm m} = q^{\sum\limits_{i = 2}^l m_{1 i}} v_{{\bm m} + \epsilon_{1, l + 1}}. \label{3.7}
\end{gather*}
The reduction to the special linear case from the general linear one can obviously be obtained by replacing equation~(\ref{3.3}) by
\begin{gather*}
q^{\nu H_i} v_{\bm m} = q^{\nu \big[\lambda_i - \lambda_{i+1} + \sum\limits_{k=1}^{i-1} (m_{ki} - m_{k,i+1}) - 2 m_{i,i+1}- \sum\limits_{k=i+2}^{l+1} (m_{ik} - m_{i+1,k})\big]} v_{\bm m}. \label{3.3x}
\end{gather*}
It is clear that $\widetilde V^\lambda$ and $\widetilde \pi^\lambda$ are inf\/inite-dimensional for the general weights $\lambda \in \gothk^*_{l+1}$. However, if all the dif\/ferences $\lambda_i - \lambda_{i+1}$, $i = 1,\ldots,l$, are non-negative integers, there is a maximal submodule, such that the respective quotient module is f\/inite-dimensional. This quotient is then denoted by $V^\lambda$ and the corresponding representation is denoted by $\pi^\lambda$.

\section[Quantum loop algebra $\uqlsllpo$ and its representations]{Quantum loop algebra $\boldsymbol{\uqlsllpo}$ and its representations}\label{s:qlars}

\subsection{Cartan--Weyl data} \label{ss:cwd}

It is convenient to introduce the sets $I = \{1, \ldots,l\}$ and $\hi = \{0, 1, \ldots,l\}$. We use notations adopted by Kac in his book \cite{Kac90}. Thus, $\lsllpo$ means the loop algebra of $\sllpo$, $\tlsllpo$ its standard extension by a one-dimensional center~$\bbC c$, and~$\hlsllpo$ the Lie algebra obtained from~$\tlsllpo$ by adding a natural derivation~$d$. The Cartan subalgebra~$\hgothh_{l+1}$ of~$\hlsllpo$ is
\begin{gather*}
\hgothh_{l+1} = \gothh_{l+1} \oplus \bbC c \oplus \bbC d.
\end{gather*}
Denote by $h_i$, $i \in I$, the generators of the standard Cartan subalgebra of $\sllpo$ considered as a~subalgebra of~$\hlsllpo$. Introducing an additional Kac--Moody generator
\begin{gather*}
h_0 = c - \sum_{i \in I} h_i
\end{gather*}
we obtain
\begin{gather*}
\hgothh_{l+1} = \biggl( \bigoplus_{i \in \hi} \bbC h_i \biggr) \oplus \bbC d.
\end{gather*}
We identify the space $\gothh^*_{l+1}$ with the subspace of $\widehat \gothh^*_{l+1}$ formed by the elements $\gamma \in \widehat \gothh^*_{l+1}$ satisfying the equations
\begin{gather*}
\langle \gamma, c \rangle = 0, \qquad \langle \gamma, d \rangle = 0.
\end{gather*}
We also denote
\begin{gather*}
\tgothh_{l+1} = \gothh_{l+1} \oplus \bbC c.
\end{gather*}
Similarly as above, we denote the generators of the standard Cartan subalgebra of $\sllpo$ considered as a subalgebra of $\tgothh_{l+1}$ by $h_i$. Then we can write
\begin{gather*}
\tgothh_{l+1} = \Bigl( \bigoplus_{i \in I} \bbC h_i \Bigr) \oplus \bbC c = \bigoplus_{i \in \widehat I} \bbC h_i.
\end{gather*}
In fact, below we use the notation $h_i$ even for the generators of the standard Cartan subalgebra of $\sllpo$ itself. This never leads to misunderstanding. We identify the space $\gothh^*_{l+1}$ with the subspace of $\widetilde \gothh^*_{l+1}$ which consists of the elements $\widetilde \gamma \in \widetilde \gothh^*_{l+1}$ subject to the condition
\begin{gather}
\langle \widetilde \gamma, c \rangle = 0.
\label{lambdac}
\end{gather}
Here and everywhere below we mark such elements by a tilde. Explicitly the identif\/ication is performed as follows. The element $\widetilde \gamma \in \widetilde \gothh^*_{l+1}$ satisfying~(\ref{lambdac}) is identif\/ied with the element $\gamma \in \gothh^*_{l+1}$ def\/ined by the equations
\begin{gather*}
\langle \gamma, h_i \rangle = \langle \widetilde \gamma, h_i \rangle, \qquad i \in I.
\end{gather*}
In the opposite direction, given an element $\gamma \in \gothh^*_{l+1}$, we identify it with the element $\widetilde \gamma \in \widetilde \gothh^*_{l+1}$ determined by the relations
\begin{gather*}
\langle \widetilde \gamma, h_0 \rangle = - \sum_{i \in I} \langle \gamma, h_i \rangle, \qquad \langle \widetilde \gamma, h_i \rangle = \langle \gamma, h_i \rangle, \quad i \in I.
\end{gather*}
It is clear that $\widetilde \gamma$ satisf\/ies (\ref{lambdac}).

The simple roots $\alpha_i \in \hgothh^*_{l+1}$, $i \in \hi$, of the Lie algebra
$\hlsllpo$ are def\/ined by the relations
\begin{gather*}
\langle \alpha_i, h_j \rangle = a_{j i}, \quad i, j \in \widehat I, \qquad
\langle \alpha_0, d \rangle = 1, \qquad \langle \alpha_i, d \rangle = 0,
\quad i \in I.
\end{gather*}
Here $a_{i j}$, $i, j \in \hi$, are the entries of the extended Cartan matrix of $\sllpo$. The full system $\widehat \triangle_+$ of positive roots of the Lie algebra $\hlsllpo$ is related to the system $\triangle_+$ of positive roots of $\sllpo$ as
\begin{gather*}
\widehat \triangle_+ = \{\gamma + n \delta \,|\, \gamma \in \triangle_+, \, n \in \bbZ_+\} \cup \{n \delta \,|\, n \in \bbN\} \cup \{(\delta - \gamma) + n \delta \,|\, \gamma \in \triangle_+, \, n \in \bbZ_+\},
\end{gather*}
where
\begin{gather*}
\delta = \sum_{i \in \hi} \alpha_i
\end{gather*}
is the minimal positive imaginary root. We note here that
\begin{gather*}
\alpha_0 = \delta - \sum_{i \in I} \alpha_i = \delta - \theta,
\end{gather*}
where $\theta$ is the highest root of $\sllpo$. The system of negative roots $\widehat \triangle_-$ is $\widehat \triangle_- = - \widehat \triangle_+$, and the full system of roots is
\begin{gather*}
\widehat \triangle = \widehat \triangle_+ \sqcup \widehat \triangle_- = \{ \gamma + n \delta \,|\, \gamma \in \triangle, \, n \in \bbZ \} \cup \{n \delta \,|\, n \in \bbZ \setminus \{0\} \}.
\end{gather*}
The set formed by the restriction of the simple roots $\alpha_i$ to $\tgothh_{l+1}$ is linearly dependent, as the restriction of $\delta$ on $\tgothh_{l+1}$ evidently vanishes. This is exactly why we pass from $\tlsllpo$ to $\hlsllpo$.

A non-degenerate symmetric bilinear form on $\hgothh_{l+1}$ is f\/ixed by the equations
\begin{gather*}
(h_i \,|\, h_j) = a^{\mathstrut}_{i j}, \qquad (h_i \,|\, d) = \delta_{i 0}, \qquad (d \,|\, d) = 0,
\end{gather*}
where $i, j \in \hi$. For the corresponding symmetric bilinear form on~$\hgothh^*_{l+1}$ one has
\begin{gather*}
(\alpha_i \,|\, \alpha_j) = a_{i j}.
\end{gather*}
This relation implies that
\begin{gather*}
(\delta \,|\, \alpha_{i j}) = 0, \qquad (\delta \,|\, \delta) = 0
\end{gather*}
for all $1 \le i < j \le l + 1$.

To def\/ine the quantum loop algebra $\uqlsllpo$, it is reasonable to start with the quantum group $\uqhlsllpo$. The latter is generated by the elements~$e_i$, $f_i$, $i \in \hi$, and $q^x$, $x \in \hgothh_{l+1}$, subject to the relations
\begin{gather}
q^0 = 1, \qquad q^{x_1} q^{x_2} = q^{x_1 + x_2}, \label{djra} \\
q^x e_i q^{-x} = q^{\langle \alpha_i, x \rangle} e_i, \qquad q^x f_i q^{-x} = q^{- \langle \alpha_i, x \rangle} f_i, \\
[e_i, f_j] = \delta_{i j} \frac{q^{h_i} - q^{- h_i}}{q^{\mathstrut} - q^{-1}}, \\
\sum_{k = 0}^{1 - a_{i j}} (-1)^k e_i^{(1 - a_{i j} - k)} e^{\mathstrut}_j e_i^{(k)} = 0, \qquad
\sum_{k = 0}^{1 - a_{i j}} (-1)^k f_i^{(1 - a_{i j} - k)} f^{\mathstrut}_j f_i^{(k)} = 0,\label{djrd}
\end{gather}
where $e_i^{(n)} = e_i^n / [n]_{q}!$, $f_i^{(n)} = f_i^n / [n]_{q}!$, and the indices $i$ and $j$ in the Serre relations (\ref{djrd}) are distinct. $\uqhlsllpo$ is a Hopf algebra with respect to the co-multiplication, antipode and co-unit def\/ined as
\begin{gather*}
\Delta(q^x) = q^x \otimes q^x, \qquad \Delta(e_i) = e_i \otimes 1 + q^{h_i} \otimes e_i, \qquad \Delta(f_i) = f_i \otimes q^{-h_i} + 1 \otimes f_i, \\
S(q^x) = q^{- x}, \qquad S(e_i) = - q^{-h_i} e_i, \qquad S(f_i) = - f_i q^{h_i}, \\
\epsilon(q^x) = 1, \qquad \epsilon(e_i) = 0, \qquad \epsilon(f_i) = 0.
\end{gather*}
The quantum group $\uqhlsllpo$ does not have any f\/inite-dimensional representations with a~nontrivial action of the element~$q^{\nu c}$~\cite{ChaPre91, ChaPre94}. In contrast, the quantum loop algebra $\uqlsllpo$ possesses, apart from the inf\/inite-dimensional representations, also nontrivial f\/inite-dimensional representations. Therefore, we proceed to the quantum loop algebra $\uqlsllpo$. First, we def\/ine the quantum group $\uqtlsllpo$ as a Hopf subalgebra of $\uqhlsllpo$ generated by the elements $e_i$, $f_i$, $i \in \hi$, and $q^x$, $x \in \tgothh_{l+1}$, with relations~(\ref{djra})--(\ref{djrd}) and the above Hopf algebra structure. Second, the quantum loop algebra $\uqlsllpo$ is def\/ined as the quotient algebra of~$\uqtlsllpo$ by the two-sided Hopf ideal generated by the elements of the form~$q^{\nu c} - 1$ with~\smash{$\nu \in \bbC^\times$}. It is convenient to treat the quantum loop algebra~$\uqlsllpo$ as a complex algebra with the same generators as~$\uqwtLsllpo$, but satisfying, additionally to rela\-tions~\mbox{(\ref{djra})--(\ref{djrd})}, also the relations
\begin{gather*}
q^{\nu c} = 1, \qquad \nu \in \bbC^\times. \label{qnuc}
\end{gather*}

For the quantum group under consideration one can also def\/ine the root vectors and construct a Poincar\'e--Birkhof\/f--Witt basis. This basis is used, in particular, to relate two realizations of the quantum loop algebra. To def\/ine the root vectors, we introduce the root lattice of~$\hlsllpo$. This is the abelian group
\begin{gather*}
\widehat Q = \bigoplus_{i \in \widehat I} \bbZ \alpha_i.
\end{gather*}
The algebra $\uqlsllpo$ becomes $\widehat Q$-graded if we assume
\begin{gather*}
e_i \in \uqlsllpo_{\alpha_i}, \qquad f_i \in \uqlsllpo_{- \alpha_i}, \qquad q^x \in \uqlsllpo_0
\end{gather*}
for any $i \in \widehat I$ and $x \in \tgothh$. Then, an element $a$ of $\uqlsllpo$ is called a root vector corresponding to a root $\gamma$ of $\hlsllpo$ if $a \in \uqlsllpo_\gamma$. The generators $e_i$ and $f_i$ are root vectors corresponding to the roots~$\alpha_i$ and~$- \alpha_i$.

Now we obtain linearly independent root vectors corresponding to the roots from $\widehat \triangle$. We use here the procedure of Khoroshkin and Tolstoy
\cite{KhoTol93, TolKho92} as the most suitable for the purpose. The root vectors, together with the elements $q^x$, $x \in \widetilde \gothh$, are the Cartan--Weyl generators of~$\uqlsllpo$.

We endow $\widehat \triangle_+$ with an order $\prec$ in the following way. First we assume that imaginary roots follow each other in any order. Then we additionally assume that
\begin{gather}
\alpha + k \delta \prec m \delta \prec (\delta - \beta) + n \delta \label{akd}
\end{gather}
for any $\alpha, \beta \in \triangle_+$ and $k, m, n \in \bbZ_+$. Finally we impose a normal order in the sense of \cite{AshSmiTol79, LezSav74} on the system of real positive roots from~$\triangle_+$. We specify this normal order as described, for example, in~\cite{MenTes15}. It is clear from~(\ref{akd}) that it is suf\/f\/icient to def\/ine the ordering separately for the roots $\alpha + k \delta$ and $(\delta - \beta) + n \delta$, where $\alpha, \beta \in
\triangle_+$. We assume that $\alpha_{i j} + r \delta \prec \alpha_{m n} + s \delta$ if $i < m$, or if $i = m$ and $r < s$, or if $i = m$, $r = s$ and $j < n$. Similarly, $(\delta - \alpha_{i j}) + r \delta \prec (\delta - \alpha_{m n}) + s \delta$ if $i > m$, or if $i = m$ and $r > s$, or if $i = m$, $r = s$ and $j < n$. The restriction of this ordering to~$\triangle_+$ gives the lexicographic ordering described in the preceding Section~\ref{s:qglrs}.

The root vectors can be def\/ined inductively. We start with the root vectors corresponding to the simple roots, which are nothing but the generators of~$\uqlsllpo$,
\begin{gather*}
e_{\delta - \theta} = e_0, \qquad e_{\alpha_i} = e_i, \qquad f_{\delta - \theta} = f_0, \qquad f_{\alpha_i} = f_i, \qquad i \in I.
\end{gather*}
As usual, a root vector corresponding to a positive root $\gamma$ is denoted by $e_\gamma$, and a root vector corresponding to a negative root $- \gamma$ is denoted by $f_\gamma$. Let a root $\gamma \in \widehat \triangle_+$ be such that $\gamma = \alpha + \beta$ for some $\alpha, \beta \in \widehat \triangle_+$. For def\/initeness, we assume that $\alpha \prec \gamma \prec \beta$, and there are no other roots $\alpha' \succ \alpha$ and $\beta' \prec \beta$ such
that $\gamma = \alpha' + \beta'$. Then, if the root vectors $e_\alpha$, $e_\beta$ and $f_\alpha$, $f_\beta$ are already def\/ined, we put~\cite{KhoTol93,TolKho92}
\begin{gather*}
e_\gamma = [e_\alpha , e_\beta]_q, \qquad f_\gamma = [f_\beta , f_\alpha]_q,\label{e1}
\end{gather*}
where the $q$-commutator $[ \ , \ ]_q$ is def\/ined by the relations
\begin{gather*}
[e_\alpha , e_\beta]_q= e_\alpha e_\beta - q^{-(\alpha | \beta)} e_\beta e_\alpha, \qquad
[f_\alpha , f_\beta]_q= f_\alpha f_\beta - q^{(\alpha | \beta)} f_\beta f_\alpha
\end{gather*}
with $( \ | \ )$ standing for the symmetric bilinear form on $\widehat \gothh^*$.

Next we def\/ine root vectors corresponding to the roots $\alpha_{i j}$ and $- \alpha_{i j}$. The root vectors $e_{\alpha_{i, i + 1}}$ and $f_{\alpha_{i, i + 1}}$ corresponding to the roots $\alpha_{i, i + 1} = \alpha_i$ and $- \alpha_{i, i + 1} = - \alpha_i$ are already given. The higher root vectors for the positive and negative composite roots can be def\/ined by the relations
\begin{gather*}
e_{\alpha_{i j}} = [e_{\alpha_{i, i + 1}}, e_{\alpha_{i + 1, j}}]_q = e_{\alpha_{i, i + 1}} e_{\alpha_{i + 1, j}} - q e_{\alpha_{i + 1, j}} e_{\alpha_{i, i + 1}}
\end{gather*}
and
\begin{gather*}
f_{\alpha_{i j}} = [f_{\alpha_{i + 1, j}}, f_{\alpha_{i, i + 1}}]_q = f_{\alpha_{i + 1, j}} f_{\alpha_{i, i + 1}} - q^{-1} f_{\alpha_{i, i + 1}} f_{\alpha_{i + 1, j}}, 
\end{gather*}
respectively. These def\/initions uniquely give
\begin{gather*}
e_{\alpha_{i j}} = [e_{\alpha_i}, \ldots [e_{\alpha_{j - 2}}, e_{\alpha_{j - 1}} ]_q \ldots ]_q,
\qquad f_{\alpha_{i j}} = [ \ldots [f_{\alpha_{j - 1}}, f_{\alpha_{j - 2}}]_q \ldots, f_{\alpha_i}]_q.
\end{gather*}
In general, we begin with the simple root $\alpha_{j - 1} = \alpha_{j - 1, j}$ and sequentially append necessary simple roots from the left to obtain the f\/inal root $\alpha_{i j}$. We can certainly start with any simple root $\alpha_k$ with $i < k < j$ and go by adding the appropriate simple roots from the left or from the right in arbitrary order. However, the resulting root vector will always be the same.

Further, we proceed to the roots of the form $\delta - \alpha_{i j}$ and $- (\delta - \alpha_{i j})$. First, we note that the root vectors $e_{\delta - \theta}$ and $f_{\delta - \theta}$ corresponding to the roots $\delta - \theta = \delta - \alpha_{1, l + 1}$ and $-(\delta - \theta) = - (\delta - \alpha_{1, l + 1})$ are already given. Then, we def\/ine inductively
\begin{gather}
e_{\delta - \alpha_{i j}} = [e_{\alpha_{i - 1, i}}, e_{\delta - \alpha_{i - 1, j}}]_q = e_{\alpha_{i - 1, i}} e_{\delta - \alpha_{i - 1, j}} - q e_{\delta - \alpha_{i - 1, j}} e_{\alpha_{i - 1, i}}, \label{edmga} \\
 f_{\delta - \alpha_{i j}} = [f_{\delta - \alpha_{i - 1, j}}, f_{\alpha_{i - 1, i}}]_q = f_{\delta - \alpha_{i - 1, j}} f_{\alpha_{i - 1, i}} - q^{-1} f_{\alpha_{i - 1, i}} f_{\delta - \alpha_{i - 1, j}} \label{edmgb}
\end{gather}
if $i > 1$, and
\begin{gather}
e_{\delta - \alpha_{1 j}} = [e_{\alpha_{j, j + 1}},
e_{\delta - \alpha_{1, j + 1}}]_q = e_{\alpha_{j, j + 1}}
e_{\delta - \alpha_{1, j + 1}} - q e_{\delta - \alpha_{1, j + 1}}
e_{\alpha_{j, j + 1}}, \label{edmgc} \\
f_{\delta - \alpha_{1 j}} = [f_{\delta - \alpha_{1, j + 1}},
f_{\alpha_{j, j + 1}}]_q = f_{\delta - \alpha_{1, j + 1}}
f_{\alpha_{j, j + 1}} - q^{-1} f_{\alpha_{j, j + 1}}
f_{\delta - \alpha_{1, j + 1}} \label{edmgd}
\end{gather}
for $j < l + 1$. The inductive rules (\ref{edmga}), (\ref{edmgb}) and (\ref{edmgc}), (\ref{edmgd}) uniquely
lead to the expressions
\begin{gather}
e_{\delta - \alpha_{i j}} = [e_{\alpha_{i - 1}} , \ldots [e_{\alpha_1} , [e_{\alpha_{j}} , \ldots [e_{\alpha_l} , e_{\delta - \theta}]_q \ldots ]_q ]_q \ldots ]_q, \label{edaij} \\
f_{\delta - \alpha_{i j}} = [\ldots [ [ \ldots [ f_{\delta - \theta} , f_{\alpha_l} ]_q , \ldots f_{\alpha_{j}} ]_q , f_{\alpha_1} ]_q , \ldots f_{\alpha_{i - 1}} ]_q. \label{fdaij}
\end{gather}
Generally speaking, we begin with the highest root $\theta$ and sequentially subtract redundant simple roots f\/irst from the right and then from the left. We can arbitrarily interchange subtractions from the left and from the right, but the result will be the same. Indeed, there is another obvious possibility to write relations (\ref{edaij}), (\ref{fdaij}), namely
\begin{gather*}
e_{\delta - \alpha_{i j}} = [e_{\alpha_{j}} , \ldots [e_{\alpha_l} ,
[e_{\alpha_{i-1}} , \ldots [e_{\alpha_1} , e_{\delta - \theta}]_q \ldots ]_q ]_q \ldots ]_q, \\
f_{\delta - \alpha_{i j}} = [\ldots [ [ \ldots [ f_{\delta - \theta} , f_{\alpha_1} ]_q
 , \ldots f_{\alpha_{i-1}} ]_q , f_{\alpha_l} ]_q , \ldots f_{\alpha_{j}} ]_q.
\end{gather*}

Finally, for $j = i + 1$, so that $\alpha_{i j}$ at the left hand side of relations (\ref{edaij}), (\ref{fdaij}) means any of the simple roots $\alpha_{i, i + 1} = \alpha_i$, $i \in I$, we obtain
\begin{gather}
e_{\delta - \alpha_1} = [ e_{\alpha_2} , [ e_{\alpha_3} , \ldots [e_{\alpha_l}
 , e_{\delta - \theta}]_q \ldots ]_q ]_q, \label{edma1} \\
e_{\delta - \alpha_i} = [e_{\alpha_{i-1}} , \ldots [e_{\alpha_1} ,
[e_{\alpha_{i+1}} , \ldots [e_{\alpha_l} , e_{\delta - \theta}]_q \ldots ]_q ]_q
\ldots ]_q, \quad i = 2, \ldots, l-1, \label{edmai} \\
e_{\delta - \alpha_l} = [e_{\alpha_{l-1}} , \ldots [e_{\alpha_2} ,
[e_{\alpha_1} , e_{\delta - \theta}]_q]_q \ldots ]_q \label{edmal}
\end{gather}
and, similarly,
\begin{gather}
f_{\delta - \alpha_1} = [ [ \ldots [ f_{\delta - \theta} , f_{\alpha_l} ]_q ,
\ldots f_{\alpha_3} ]_q , f_{\alpha_2} ]_q, \label{fdma1} \\
f_{\delta - \alpha_i} = [\ldots [ [ \ldots [ f_{\delta - \theta} , f_{\alpha_l} ]_q
 , \ldots f_{\alpha_{i+1}} ]_q , f_{\alpha_1} ]_q , \ldots f_{\alpha_{i-1}} ]_q,
\quad i = 2, \ldots, l - 1, \label{fdmai} \\
f_{\delta - \alpha_l} = [ \ldots [ [ f_{\delta - \theta} , f_{\alpha_1} ]_q ,
f_{\alpha_2} ]_q , \ldots f_{\alpha_{l - 1}} ]_q. \label{fdmal}
\end{gather}
It is clear that (\ref{edmai}) and (\ref{fdmai}) can also be written equivalently as
\begin{align*}
e_{\delta - \alpha_i} & = [e_{\alpha_{i+1}} , \ldots [e_{\alpha_l} ,
[e_{\alpha_{i-1}} , \ldots [e_{\alpha_1} , e_{\delta - \theta}]_q \ldots ]_q ]_q
\ldots ]_q, \\
f_{\delta - \alpha_i} & = [\ldots [ [ \ldots [ f_{\delta - \theta} , f_{\alpha_1} ]_q
 , \ldots f_{\alpha_{i-1}} ]_q , f_{\alpha_l} ]_q , \ldots f_{\alpha_{i+1}} ]_q.
\end{align*}

When the root vectors corresponding to all the roots $\alpha_{i j}$ and
$\delta - \alpha_{i j}$ with $\alpha_{i j} \in \triangle_+$ are def\/ined,
we can continue by adding imaginary roots $n \delta$. The root vectors
corresponding to the imaginary roots are additionally labelled by the
positive roots $\gamma \in \triangle_+$ of $\sllpo$ and are given by
the relations
\begin{gather*}
e'_{\delta, \gamma} = [e_\gamma , e_{\delta - \gamma}]_q,
\qquad f'_{\delta, \gamma} = [f_{\delta - \gamma} , f_{\gamma}]_q.
\label{epd}
\end{gather*}
The remaining higher root vectors are def\/ined iteratively by \cite{KhoTol93,TolKho92}
\begin{gather*}
e_{\gamma + n \delta} = ([2]_q)^{-1}
[e_{\gamma + (n - 1)\delta} , e'_{\delta, \gamma}]_q, \qquad
f_{\gamma + n \delta} = ([2]_q)^{-1}
[f'_{\delta, \gamma} , f_{\gamma + (n - 1)\delta}]_q, \\
e_{(\delta - \gamma) + n \delta} = ([2]_q)^{-1}
[e'_{\delta, \gamma} , e_{(\delta - \gamma) + (n - 1)\delta}]_q, \\
f_{(\delta - \gamma) + n \delta} = ([2]_q)^{-1}
[f_{(\delta - \gamma) + (n - 1)\delta} , f'_{\delta, \gamma}]_q, \\
e'_{n \delta, \gamma} = [e_{\gamma + (n - 1)\delta} , e_{\delta - \gamma}]_q,
\qquad f'_{n \delta, \gamma} = [f_{\delta - \gamma} , f_{\gamma + (n - 1)\delta}]_q,
\end{gather*}
where we use that $(\alpha_{i j} | \alpha_{i j}) = 2$ for all $i$ and $j$. It is worth to note that, among all imaginary root vectors $e'_{n\delta, \gamma}$ and $f'_{n\delta, \gamma}$ only the root vectors $e'_{n \delta, \alpha_i}$ and $f'_{n \delta, \alpha_i}$, $i \in I$, are independent and required for the construction of the Poincar\'e--Birkhof\/f--Witt basis.

Besides, there is another set of useful root vectors introduced by the functional equations
\begin{gather}
- \kappa_q e_{\delta, \gamma}(u) = \log(1 - \kappa_q e'_{\delta, \gamma}(u)), \label{edg} \\
\kappa_q f_{\delta, \gamma}(u^{-1}) = \log(1 + \kappa_q f'_{\delta, \gamma}(u^{-1})), \label{fdg}
\end{gather}
where the generating functions
\begin{alignat*}{3}
& e'_{\delta, \gamma}(u) = \sum_{n = 1}^\infty e'_{n \delta, \gamma} u^n,\qquad && e_{\delta, \gamma}(u) = \sum_{n = 1}^\infty e_{n \delta, \gamma} u^n, & \\
& f'_{\delta, \gamma}(u^{-1}) = \sum_{n = 1}^\infty f'_{n \delta, \gamma} u^{- n},\qquad && f_{\delta, \gamma}(u^{-1}) = \sum_{n = 1}^\infty f_{n \delta, \gamma} u^{- n}&
\end{alignat*}
are def\/ined as formal power series. The unprimed imaginary root vectors arise, for example, in formulas for the universal $R$-matrix of quantum af\/f\/ine algebras \cite{KhoTol93, TolKho92}.

\subsection{Drinfeld's second realization} \label{ss:2dr}

Drinfeld realized $\uqlsllpo$ also in a dif\/ferent way \cite{Dri87, Dri88}, as an algebra generated by $\xi^\pm_{i, n}$, $i \in I$, $n \in \bbZ$, $q^x$, $x \in \gothh$, and $\chi_{i, n}$, $i \in I$, $n \in \bbZ \setminus \{0\}$. These generators satisfy the def\/ining relations
\begin{gather*}
q^0 = 1, \qquad q^{x_1} q^{x_2} = q^{x_1 + x_2}, \\
[\chi^{\mathstrut}_{i, n}, \chi^{\mathstrut}_{j, m}] = 0, \qquad q^x \chi_{j, n} = \chi_{j, n} q^x, \\
q^x \xi^\pm_{i, n} q^{- x} = q^{\pm \langle \alpha_i, x \rangle} \xi^\pm_{i, n}, \qquad [\chi^{\mathstrut}_{i, n}, \xi^\pm_{j, m}] = \pm \frac{1}{n}
[n a_{i j}]^{\mathstrut}_{q} \xi^\pm_{j, n + m}, \\
\xi^\pm_{i, n + 1} \xi^\pm_{j, m} - q^{\pm a_{i j}} \xi^\pm_{j, m} \xi^\pm_{i, n + 1} = q^{\pm a_{i j}} \xi^\pm_{i, n}
\xi^\pm_{j, m + 1} - \xi^\pm_{j, m + 1} \xi^\pm_{i, n}, \\
[\xi^+_{i, n}, \xi^-_{j, m}] = \delta_{i j} \frac{\phi^+_{i, n + m} - \phi^-_{i, n + m}}{q^{\mathstrut} - q^{-1}}.
\end{gather*}
Besides, there are the Serre relations. However, their explicit form is not relevant, and so, we do not put them here. In the above relations, $a_{i j}$ are the entries of the Cartan matrix of $\gothsl_{l+1}$. The quantities $\phi^\pm_{i, n}$, $i \in I$, $n \in \bbZ$, are given by the formal power series
\begin{gather}
\sum_{n = 0}^\infty \phi^\pm_{i, \pm n} u^{\pm n} = q^{\pm h_i} \exp \left( \pm \kappa_q \sum_{n = 1}^\infty \chi_{i, \pm n} u^{\pm n} \right), \label{phipm}
\end{gather}
where the conditions
\begin{gather*}
\phi^+_{i, n} = 0, \quad n < 0, \qquad \phi^-_{i, n} = 0, \quad n > 0
\end{gather*}
are assumed.

There is an isomorphism of the two realizations of the quantum loop algebras. In the case under consideration, the generators of the Drinfeld's second realization are connected with the Cartan--Weyl generators as follows \cite{KhoTol93, KhoTol94}. The generators $q^x$ in the Drinfeld--Jimbo's and Drinfeld's second realizations are the same, with an important exception that in the f\/irst case $x \in \widetilde \gothh$, and in the second case $x \in \gothh \subset \widetilde \gothh$. For the generators $\xi^\pm_{i, n}$ and $\chi_{i, n}$ of the Drinfeld's second realization one has explicitly
\begin{gather}
\xi^+_{i, n} = \begin{cases}
(-1)^{n i} e_{\alpha_i + n \delta}, & n \ge 0, \\
-(-1)^{n i} q^{-h_i} f_{(\delta - \alpha_i) - (n + 1)\delta}, & n < 0,
\end{cases} \label{ksipin} \\
\xi^-_{i, n} = \begin{cases}
-(-1)^{(n + 1) i} e_{(\delta - \alpha_i) + (n - 1) \delta} q^{h_i}, & n > 0, \\
(-1)^{n i} f_{\alpha_i - n \delta}, & n \le 0,
\end{cases} \label{ksimin} \\
\chi_{i, n} = \begin{cases}
-(-1)^{n i} e_{n \delta, \alpha_i}, & n > 0, \\
-(-1)^{n i} f_{- n\delta, \alpha_i}, & n < 0.
\end{cases} \label{chiin}
\end{gather}
As follows from (\ref{edg}), (\ref{fdg}), (\ref{phipm}) and (\ref{chiin}),
\begin{gather*}
\phi^+_{i, n} = \begin{cases}
-(-1)^{n i} \kappa_q q^{h_i} e'_{n \delta, \alpha_i}, & n > 0, \\
q^{h_i}, & n = 0,
\end{cases} \qquad
\phi^-_{i, n} = \begin{cases}
q^{-h_i}, & n = 0 ,\\
(-1)^{n i} \kappa_q q^{-h_i} f'_{- n \delta, \alpha_i}, & n < 0.
\end{cases}
\end{gather*}
Introducing the generating functions $\phi^+_i(u)$ and $\phi^-_i(u)$ by the formal power series
\begin{gather*}
\phi^+_i(u) = \sum_{n = 0}^\infty \phi^+_{i, n} u^n, \qquad \phi^-_i(u^{-1}) = \sum_{n = 0}^\infty \phi^-_{i, -n} u^{- n},
\end{gather*}
we obtain
\begin{gather}
\phi^+_i(u) = q^{h_i} \big(1 - \kappa_q e'_{\delta, \alpha_i}\big((-1)^i u\big)\big), \label{phipiu} \\
\phi^-_i(u^{-1}) = q^{-h_i} \big(1 + \kappa_q f'_{\delta, \alpha_i}\big((-1)^i u^{-1}\big)\big). \label{phimiu}
\end{gather}

We refer also to \cite{Beck94}, where this isomorphism between two realizations of the untwisted quantum loop algebra is established by means of a dif\/ferent approach. Besides, in \cite{Dam12} for more general case of twisted af\/f\/ine quantum algebras it was shown that the relation between the two realizations of the quantum group, def\/ined as a $\bbC(q)$-algebra, is given by a surjective homomorphism from the Drinfeld's second realization to the Drinfeld--Jimbo's realization, and later, in~\cite{Dam15}, this surjective homomorphism was shown to be injective, thus
proving the isomorphism between the two realizations.

\subsection{Jimbo's homomorphism} \label{ss:jh}

Highest weight representations of the quantum loop algebra $\uqlsllpo$ are based on the eva\-luation homomorphism $\varepsilon$ from $\uqlsllpo$ to~$\uqgllpo$~\cite{Jim86a}. It is def\/ined by the relations
\begin{alignat*}{3}
& \varepsilon(q^{\nu h_0}) = q^{\nu (K_{l+1} - K_1)}, \qquad &&
 \varepsilon(q^{\nu h_i}) = q^{\nu (K_{i} - K_{i+1})},& \\
& \varepsilon(e_0) = F_{1, l+1} q^{K_1 + K_{l+1}}, \qquad &&
 \varepsilon(e_i) = E_{i, i+1}, & \\
& \varepsilon(f_0) = E_{1, l+1} q^{-K_1 - K_{l+1}}, \qquad &&
 \varepsilon(f_i) = F_{i, i+1},&
\end{alignat*}
where $i$ takes all integer values from $1$ to $l$. Thus, if $\pi$ is a representation of~$\uqgllpo$, then the composition $\pi \circ \varepsilon$ gives a representation of~$\uqlsllpo$.

In quantum integrable systems, one considers families of representations parameterized by the so-called spectral parameters. We introduce a spectral parameter by means of the mappings $\Gamma_\zeta \colon\uqlsllpo \to \uqlsllpo$, $\zeta \in \bbC^\times$, def\/ined explicitly by the following action on the generators:
\begin{gather*}
\Gamma_\zeta(q^x) = q^x, \qquad \Gamma_\zeta(e_i) = \zeta^{s_i} e_i, \qquad \Gamma_\zeta(f_i) = \zeta^{-s_i} f_i.
\end{gather*}
Here, $s_i$ are arbitrary integers, and it is convenient to denote their total sum by~$s$. Further, given any representation~$\varphi$ of~$\uqlsllpo$, we def\/ine the corresponding family $\varphi_\zeta$ of representations as
\begin{gather*}
\varphi_\zeta = \varphi \circ \Gamma_\zeta.
\end{gather*}
We are interested in the representations $(\widetilde \varphi^\lambda)_\zeta$ and $(\varphi^\lambda)_\zeta$ related to inf\/inite- and f\/inite-di\-men\-si\-on\-al representations $\widetilde \pi^\lambda$ and~$\pi^\lambda$ of~$\uqgllpo$. They are def\/ined as
\begin{gather*}
\big(\widetilde \varphi{}^\lambda\big)_\zeta = \widetilde \pi^\lambda \circ \varepsilon \circ \Gamma_\zeta,
\qquad \big(\varphi^\lambda\big)_\zeta = \pi^\lambda \circ \varepsilon \circ \Gamma_\zeta.
\end{gather*}
Slightly abusing notation we denote the corresponding $\uqlsllpo$-modules by $\widetilde V^\lambda$ and $V^\lambda$. The def\/ining relations for these modules were obtained in \cite{NirRaz16b} and are as follows:
\begin{gather*}
q^{\nu h_0} v_{\bm m} = q^{\nu \big[\lambda_{l + 1} - \lambda_1 + \sum\limits_{i = 2}^l (m_{1 i} + m_{i, l + 1}) + 2 m_{1, l + 1}\big]} v_{\bm m},\\
q^{\nu h_i} v_{\bm m} = q^{\nu \big[\lambda_i - \lambda_{i + 1} + \sum\limits_{k = 1}^{i - 1} (m_{k i} - m_{k, i + 1}) - 2 m_{i, i+1} - \sum\limits_{k = i + 2}^{l + 1} (m_{i k} - m_{i + 1, k})\big]} v_{\bm m}, \\
e_0 v_{\bm m} = \zeta^{s_0} q^{\lambda_1 + \lambda_{l + 1} + \sum\limits_{i = 2}^l m_{i, l+1}} v_{{\bm m} + \epsilon_{1, l + 1}}, \\
e_i v_{\bm m} = \zeta^{s_i} \bigg[\lambda_i - \lambda_{i+1} - \sum_{j = i + 2}^{l + 1} (m_{i j} - m_{i + 1, j}) - m_{i, i + 1} + 1\bigg]_q [m_{i, i + 1}]_q v_{{\bm m} - \epsilon_{i, i + 1}} \notag \\
 \hphantom{e_i v_{\bm m} =}{} + \zeta^{s_i} q^{\lambda_i - \lambda_{i + 1} - 2 m_{i, i + 1}
- \sum\limits_{j = i + 2}^{l + 1} (m_{i j} - m_{i + 1, j})} \sum_{j = 1}^{i - 1} q^{\sum\limits_{k = j + 1}^{i - 1} (m_{k i} - m_{k, i + 1})} [m_{j, i + 1}]_q v_{{\bm m} - \epsilon_{j, i + 1} + \epsilon_{j i}}\notag \\
\hphantom{e_i v_{\bm m} =}{}
 - \zeta^{s_i} \sum_{j = i + 2}^{l + 1} q^{- \lambda_i + \lambda_{i + 1} - 2 + \sum\limits_{k = j}^{l + 1} (m_{i k} - m_{i + 1, k})} [m_{i j}]_q v_{{\bm m} - \epsilon_{i j} + \epsilon_{i + 1, j}}, \\
f_i v_{\bm m} = \zeta^{-s_i} q^{- \sum\limits_{j = 1}^{i - 1} (m_{j i} - m_{j, i + 1})}v_{{\bm m} + \epsilon_{i, i + 1}} + \zeta^{-s_i} \sum_{j=1}^{i-1}q^{- \sum\limits_{k = 1}^{j-1} (m_{k i} - m_{k, i + 1})} [m_{ji}]_q v_{{\bm m} - \epsilon_{j i} + \epsilon_{j, i + 1}},
\end{gather*}
where $i \in I$. To complete the def\/ining $\uqlsllpo$-module relations, one also needs an expression for $f_0 v_{\bm m}$, see~\cite{NirRaz16b}, but its explicit form is not used here and we omit it.

Twisting $(\widetilde \varphi{}^\lambda)_\zeta$ and $(\varphi{}^\lambda)_\zeta$ by the automorphisms of $\uqlsllpo$, we can construct more representations of this quantum loop algebra. There are two automorphisms which can be used for the purpose. They are def\/ined by the relations
\begin{gather*}
\sigma(q^{\nu h_i}) = q^{\nu h_{i + 1}}, \qquad \sigma(e_i) = e_{i + 1}, \qquad \sigma(f_i) = f_{i + 1}, \qquad i \in \hi,
\end{gather*}
where we use the identif\/ication $q^{\nu h_{l + 1}} = q^{\nu h_0}$, $e_{l + 1} = e_0$, $f_{l + 1} = f_0$, and
\begin{gather*}
\tau(q^{h_0}) = q^{h_0}, \qquad \tau(q^{h_i}) = q^{h_{l - i + 1}}, \qquad i \in I, \\
\tau(e_0) = e_0, \qquad \tau(e_i) = e_{l - i + 1}, \qquad \tau(f_0) = f_0, \qquad \tau(f_i) = f_{l - i + 1}, \qquad i \in I.
\end{gather*}
Here we have $\sigma^{l + 1} = \id$ and $\tau^2 = \id$. We note that the transfer operators related to the twisting of $(\widetilde \varphi{}^\lambda)_\zeta$ and $(\varphi{}^\lambda)_\zeta$ by any powers of $\sigma$ dif\/fer from each other only by permutations of the components of the highest weight $\lambda \in \gothk_{l+1}^*$, see, for example, \cite{BooGoeKluNirRaz14b}. However, considering representations of the Borel subalgebras of $\uqlsllpo$, we can use the automorphism $\sigma$ to obtain new interesting representations. Also the twisting of $(\widetilde \varphi{}^\lambda)_\zeta$ and $(\varphi{}^\lambda)_\zeta$ by $\tau$ leads to actually dif\/ferent representations of~$\uqlsllpo$ and its Borel subalgebras.

\subsection[$\uqbp$-modules]{$\boldsymbol{\uqbp}$-modules} \label{ss:bpm}

The quantum loop algebra $\uqlsllpo$ has two Borel subalgebras, the positive and the negative ones, denoted by $\uqbp$ and $\uqbm$, respectively. Representations of these standard Borel subalgebras are what is usually required for the application in quantum integrable systems. In terms of the generators of the Drinfeld--Jimbo's realization of the quantum loop algebra, the Borel subalgebras are def\/ined in the following simple way. The positive Borel subalgebra is the subalgebra generated by $e_i$, $i \in \hi$, and $q^x$, $x \in \tgothh$, and the negative Borel subalgebra is the subalgebra generated by $f_i$, $i \in \hi$, and~$q^x$, $x \in \tgothh$. It is important that the Borel subalgebras are Hopf subalgebras of~$\uqlsllpo$. The description of $\uqbp$ and $\uqbm$ in terms of the generators of the Drinfeld's second realization of $\uqlsllpo$ is more intricate. Based on~\mbox{(\ref{ksipin})--(\ref{chiin})}, we note that $\uqbp$ contains the Drinfeld generators $\xi^+_{i, n}$, $\xi^-_{i, m}$, $\chi_{i, m}$ with $i \in I$, $n \geq 0$ and $m > 0$, while $\uqbm$ contains the Drinfeld generators $\xi^-_{i, n}$, $\xi^+_{i, m}$, $\chi_{i, m}$ with $i \in I$, $n \leq 0$ and $m < 0$. Since the positive and negative Borel subalgebras are related by the quantum Chevalley involution, we restrict ourselves to the consideration of~$\uqbp$ only.

Restricting any representation of $\uqlsllpo$ to $\uqbp$ one comes to a~representation of~$\uqbp$. In particular, one can consider the restriction
of~$(\widetilde \varphi^\lambda)_\zeta$ and $(\varphi^\lambda)_\zeta$. The correspon\-ding $\uqbp$-module relations are obtained by singling out the
expressions for~$q^{\nu h_i} v_{\bm{m}}$ and~$e_i v_{\bm{m}}$, $i \in \hi$, from the $\uqlsllpo$-module relations in Section~\ref{ss:jh}. The representations of type~$\varphi$ introduced in Section~\ref{s:qg}, specif\/ied later as $(\widetilde \varphi^\lambda)_\zeta$ and $(\varphi^\lambda)_\zeta$, are used for the construction of the transfer operators. The representations of type $\rho$ from Section~\ref{s:qg} used for the construction of the $Q$-operators are very dif\/ferent. For the quantum integrable systems related to~$\uqlsllpo$ the desirable representations of type $\rho$ for the $Q$-operators can be obtained from $(\widetilde \varphi^\lambda)_\zeta$ as submo\-du\-les of certain degenerations, sending each dif\/ference $\lambda_i - \lambda_{i + 1}$, $i \in I$, to positive or negative inf\/inity, see, e.g.,~\cite{BazHibKho02} and~\cite{BooGoeKluNirRaz14b, NirRaz16a} for the particular cases $l = 1$ and $l = 2$.

The general case with arbitrary $l$ was considered in \cite{NirRaz16b}. There, it was shown, in particular, that the relations
\begin{gather}
q^{\nu h_0} v_{\bm m} = q^{\nu \big(2 m_{1} + \sum\limits_{j = 2}^l m_j\big)} v_{\bm m}, \label{5.4b} \\
q^{\nu h_i} v_{\bm m} = q^{\nu (m_{i + 1} - m_i)} v_{\bm m}, \qquad i = 1, \ldots, l - 1, \label{5.5b} \\
q^{\nu h_l} v_{\bm m} = q^{- \nu \big(2 m_l + \sum\limits_{i = 1}^{l - 1} m_i\big)} v_{\bm m}, \label{5.6b} \\
e_0 v_{\bm m} = q^{\sum_{j = 2}^l m_j} v_{{\bm m} + \epsilon_{1}}, \label{5.7b} \\
e_i v_{\bm m} = - q^{m_i - m_{i + 1} - 1} [m_i]_q v_{{\bm m} - \epsilon_i + \epsilon_{i + 1}}, \qquad i = 1, \ldots, l - 1,\label{5.8b} \\
e_l v_{\bm m} = - \kappa_q^{-1} q^{m_{l}} [m_{l}]_q v_{{\bm m} - \epsilon_{l}}, \label{5.9b}
\end{gather}
where $\bm m$ denotes the $l$-tuple of nonnegative integers $(m_1, \ldots, m_l)$, and ${\bm m} + \nu \epsilon_i$ means the respective shift of $m_i$, def\/ine an irreducible representation of~$\uqbp$. Actually, these are def\/ining relations for a submodule of a degeneration of a shifted $\uqbp$-module \cite{NirRaz16b}. Comparing these relations with the def\/ining relations for the representation $(\widetilde\varphi^\lambda)_\zeta$
in Section \ref{ss:jh}, we can see the limit relation between the universal transfer operator associated with $(\widetilde\varphi^\lambda)_\zeta$ and the universal $Q$-operator associated with the representation given by (\ref{5.4b})--(\ref{5.9b}). Such limit relations between the universal transfer and $Q$-operators for $l = 1$ and $l = 2$ were established in~\cite{BooGoeKluNirRaz13, NirRaz16a} and~\cite{BooGoeKluNirRaz14b}, respectively.

Relations (\ref{5.4b})--(\ref{5.9b}) def\/ine a representation of type $\rho$ described in Section~\ref{s:qg}, that is, this representation and its twisting by the automorphisms $\sigma$ and $\tau$ are exactly what we need for the construction of representations for the $Q$-operators.

\section[Highest $\ell$-weight representations]{Highest $\boldsymbol{\ell}$-weight representations} \label{s:hlwr}

\subsection[Rational $\ell$-weights]{Rational $\boldsymbol{\ell}$-weights} \label{ss:rlws}

Here we consider $\uqlsllpo$-modules in the category $\calO$ only. For the def\/inition of this category, we refer to the original papers~\cite{Her04, Her07}, and also to the later paper \cite{MukYou14} as the most appropriate for our purposes. A very useful tool to analyze these modules is the notion of $\ell$-weights and $\ell$-weight vectors, see, for example, the papers \cite{FeiJimMiwMuk16, FreRes99, MukYou14}. One def\/ines an $\ell$-weight $\bm \Psi$ as a triple
\begin{gather}
\bm \Psi = (\lambda, \bm \Psi^+, \bm \Psi^-), \label{pppm}
\end{gather}
where $\lambda \in \gothh^*$, $\bm \Psi^+$ and $\bm \Psi^-$ are $l$-tuples
\begin{gather*}
\bm \Psi^+ = (\Psi^+_i(u))_{i \in I}, \qquad \bm \Psi^- = \big(\Psi^-_i\big(u^{-1}\big)\big)_{i \in I}
\end{gather*}
of formal series
\begin{gather*}
\Psi_i^+(u) = \sum_{n \in \bbZ_+} \Psi^+_{i, n} u^n \in \bbC[[u]], \qquad \Psi_i^+(u^{-1}) = \sum_{n \in \bbZ_+} \Psi^+_{i, n} u^{-n} \in \bbC\big[\big[u^{-1}\big]\big],
\end{gather*}
such that
\begin{gather}
\Psi^+_{i, 0} = q^{\langle \lambda, h_i \rangle}, \qquad \Psi^-_{i, 0} = q^{- \langle \lambda, h_i \rangle}. \label{ppzpmz}
\end{gather}
For an $\ell$-weight vector $v$ of $\ell$-weight $\bm \Psi$ we then have
\begin{gather*}
q^x v = q^{\langle \tilde \lambda, x \rangle} v
\end{gather*}
for any $x \in \tgothh$, and
\begin{gather*}
\phi^+_{i, n} v = \Psi^+_{i, n} v, \qquad \phi^-_{i, -n} v = \Psi^-_{i, -n} v, \qquad i \in I, \quad n \in \bbZ_+,
\end{gather*}
or, equivalently,
\begin{gather*}
\phi^+_i(u) v = \Psi^+_i(u) v, \qquad \phi^-_i\big(u^{-1}\big) v = \Psi^-_i\big(u^{-1}\big) v, \qquad i \in I.
\end{gather*}
A $\uqlsllpo$-module $V$ in the category $\calO$ is called a highest $\ell$-weight module with highest $\ell$-weight~$\bm \Psi$, if there exists an $\ell$-weight vector~$v \in V$ of $\ell$-weight~$\bm \Psi$, such that
\begin{gather*}
\xi^+_{i, n} v = 0, \qquad i \in I, \quad n \in \bbZ,
\end{gather*}
and
\begin{gather*}
V = \uqlsllpo v.
\end{gather*}
Up to a scalar factor, such a vector $v$ is determined uniquely. It is called the highest $\ell$-weight vector of $V$.

If for some non-negative integers $p_i$, $i \in I$, and complex numbers $a_{i k}$, $b_{i k}$, $i \in I$, $0 \le k \le p_i$, one has
\begin{gather}
\Psi^+_i(u) = \frac{a_{i p_i} u^{p_i} + a_{i, p_i - 1} u^{p_i - 1} + \cdots
+ a_{i 0}}{b_{i p_i} u^{p_i} + b_{i, p_i - 1} u^{p_i - 1} + \cdots + b_{i 0}}, \label{psipi} \\
 \Psi^-_i(u^{-1}) = \frac{a_{i p_i} + a_{i, p_i - 1} u^{- 1} + \cdots + a_{i 0}
u^{- p_i}}{b_{i p_i} + b_{i, p_i - 1} u^{- 1} + \cdots + b_{i 0} u^{- p_i}},\label{psimi}
\end{gather}
then one says that the corresponding $\ell$-weight $\bm \Psi$ is rational. The numbers $a_{i p_i}$, $a_{i 0}$, $b_{i p_i}$, $b_{i 0}$ must be nonzero, such that
\begin{gather*}
\frac{a_{i 0}}{b_{i 0}} = q^{\langle \lambda, h_i \rangle}, \qquad \frac{a_{i p}}{b_{i p}} = q^{-\langle \lambda, h_i \rangle}.
\end{gather*}
These equations are equivalent to (\ref{ppzpmz}).

All $\ell$-weights of a $\uqlsllpo$-module in the category $\calO$ are rational, see~\cite{MukYou14} and references therein. In fact, for any rational $\ell$-weight $\bm \Psi$ there is a simple $\uqlsllpo$-module $L(\bm \Psi)$ with highest $\ell$-weight $\bm \Psi$. Any simple $\uqlsllpo$-module in the category $\calO$ is isomorphic to $L(\bm \Psi)$ for some rational $\ell$-weight $\bm \Psi$. Thus, there is a one-to-one correspondence between the rational $\ell$-weights and the equivalence classes of the simple $\uqlsllpo$-modules in the category $\calO$.

In general, the rational $\ell$-weights, given explicitly by (\ref{psipi}), (\ref{psimi}), correspond to in\-f\/i\-ni\-te-\-dimensional $\uqlsllpo$-modules. For the f\/inite-dimensional modules they have a special form, see \cite[Proposition 1]{FreRes99}.

One def\/ines the product of $\ell$-weights $\bm \Psi_1 = (\lambda_1, \bm \Psi^+_1, \bm \Psi^-_1)$ and $\bm \Psi_2 = (\lambda_2, \bm \Psi^+_2, \bm \Psi^-_2)$ as the triple
\begin{gather*}
\bm \Psi_1 \bm \Psi_2 = \big(\lambda_1 + \lambda_2, \bm \Psi^+_1 \bm \Psi^+_2, \bm \Psi^-_1 \bm \Psi^-_2\big),
\end{gather*}
where
\begin{gather*}
\bm \Psi^+_1 \bm \Psi^+_2 = \big(\Psi^+_{1 i}(u) \Psi^+_{2 i}(u)\big)_{i \in I} \qquad
\bm \Psi^-_1 \bm \Psi^-_2 = \big(\Psi^-_{1 i}\big(u^{-1}\big) \Psi^-_{2 i}\big(u^{-1}\big)\big)_{i \in I}.
\end{gather*}
Given rational $\ell$-weights $\bm \Psi_1$ and $\bm \Psi_2$, the submodule of the tensor product $L(\bm \Psi_1) \otimes L(\bm \Psi_2)$ generated by the tensor product of the highest $\ell$-weight vectors is a highest $\ell$-weight $\uqlsllpo$-module with highest $\ell$-weight $\bm \Psi_1 \bm \Psi_2$. In particular, $L(\bm \Psi_1 \bm \Psi_2)$ is a~subquotient of $L(\bm \Psi_1) \otimes L(\bm \Psi_2)$, see~\cite{MukYou14} and references therein. As in~\cite{BooGoeKluNirRaz17}, we denote such subquotient as $L(\bm \Psi_1) \ovotimes L(\bm \Psi_2)$.
Note that the operation $\ovotimes$ is associative.

In the case of the Borel subalgebra $\uqbp$ we are left with only two f\/irst components of the triple (\ref{pppm}), and we def\/ine an $\ell$-weight $\bm \Psi^+$ as a pair
\begin{gather*}
\bm \Psi = \big(\lambda, \bm \Psi^+\big),
\end{gather*}
where $\lambda \in \gothh^*$ and $\bm \Psi^+$ is an $l$-tuple
\begin{gather*}
\bm \Psi^+ = (\Psi^+_i(u))_{i \in I}
\end{gather*}
of formal series
\begin{gather*}
\Psi_i^+(u) = \sum_{n \in \bbZ_+} \Psi^+_{i, n} u^n \in \bbC[[u]],
\end{gather*}
such that
\begin{gather*}
\Psi^+_{i, 0} = q^{\langle \lambda, h_i \rangle}.
\end{gather*}
For an $\ell$-weight vector $v$ of $\ell$-weight $\bm \Psi$ one has
\begin{gather*}
q^x v = q^{\langle \tilde \lambda, x \rangle} v
\end{gather*}
for any $x \in \tgothh$, and
\begin{gather*}
\phi^+_{i, n} v = \Psi^+_{i, n} v, \qquad i \in I, \quad n \in \bbZ_+,
\end{gather*}
or, equivalently,
\begin{gather*}
\phi^+_i(u) v = \Psi^+_i(u) v, \qquad i \in I.
\end{gather*}

A $\uqbp$-module $W$ in the category $\calO$ is called a highest $\ell$-weight module with highest $\ell$-weight $\bm \Psi$ if there exists an $\ell$-weight vector $v \in W$ of $\ell$-weight $\bm \Psi$, such that
\begin{gather*}
\xi^+_{i, n} v = 0, \qquad i \in I, \quad n \in \bbZ_+,
\end{gather*}
and
\begin{gather*}
W = \uqbp v.
\end{gather*}
Such a vector $v$ is unique up to a scalar factor, and it is called the highest $\ell$-weight vector of $W$.

An $\ell$-weight $\bm \Psi$ of a $\uqbp$-module is said to be rational, if for some non-negative integers $p_i$, $q_i$, $i \in I$, and complex numbers $a_{i r}$, $b_{i s}$,
$i \in I$, $0 \le r \le p_i$, $0 \le s \le q_i$, one has
\begin{gather}
\Psi^+_i(u) = \frac{a_{i p_i} u^{p_i} + a_{i, p_i - 1} u^{p_i - 1}
+ \cdots + a_{i 0}}{b_{i q_i} u^{q_i} + b_{i, q_i - 1} u^{q_i - 1}
+ \cdots + b_{i 0}}, \label{psipbp}
\end{gather}
where
\begin{gather*}
\frac{a_{i 0}}{b_{i 0}} = q^{\langle \lambda, h_i \rangle}.
\end{gather*}

All $\ell$-weights of a $\uqbp$-module in the category $\calO$ are rational, see \cite{HerJim12} and references therein. For any rational $\ell$-weight $\bm \Psi$
there is a simple $\uqbp$-module $L(\bm \Psi)$ with highest $\ell$-weight $\bm \Psi$.
This module is unique up to isomorphism. Any simple $\uqbp$-module is isomorphic to
$L(\bm \Psi)$ for some $\ell$-weight $\bm \Psi$. Thus, there is a one-to-one
correspondence between the rational $\ell$-weights and the equivalence classes
of the simple $\uqbp$-modules in the category~$\calO$.

Similarly as in the case of $\uqlsllpo$-modules, the general rational $\ell$-weights (\ref{psipbp}) correspond to inf\/inite-dimensional $\uqbp$-modules. For the description of f\/inite-dimensional modules we refer to \cite[Remark 3.11]{FreHer15}.

Given rational $\ell$-weights $\bm \Psi_1$ and $\bm \Psi_2$, the submodule of the tensor product $L(\bm \Psi_1) \otimes L(\bm \Psi_2)$ generated by the tensor product of the highest $\ell$-weight vectors is a highest $\ell$-weight $\uqbp$-module with highest $\ell$-weight
$\bm \Psi_1 \bm \Psi_2$. In particular, $L(\bm \Psi_1 \bm \Psi_2)$ is a subquotient of $L(\bm \Psi_1) \otimes L(\bm \Psi_2)$ denoted as $L(\bm \Psi_1) \ovotimes L(\bm \Psi_2)$.

We have already noted in Section \ref{ss:cwd} a special role of the higher root vectors~$e'_{n \delta, \alpha_i}$ and~$f'_{n \delta, \alpha_i}$. Besides, we see from equations~(\ref{phipiu}), (\ref{phimiu}) and the def\/inition of the highest $\ell$-weight representations that only the root vectors~$e'_{n \delta, \alpha_i}$, $f'_{n \delta, \alpha_i}$ are used to determine the highest $\ell$-weight vectors and highest $\ell$-weights. Relative to the Borel subalgebra~$\uqbp$, it means that only the root vectors $e'_{n \delta, \alpha_i}$ are used for the corresponding highest $\ell$-weight vectors and highest $\ell$-weights.

\subsection[Prefundamental and $q$-oscillator representations]{Prefundamental and $\boldsymbol{q}$-oscillator representations} \label{ss:examples}

The f\/irst example of the highest $\ell$-weight representations of the Borel subalgebras is given by the prefundamental representations~\cite{HerJim12}. They are def\/ined by simple highest $\ell$-weight $\uqbp$-modules $L^\pm_{i, a}$ with the highest $\ell$-weights $(\lambda^\ms_{i, a}, (\bm \Psi^\pm_{i, a})^+)$ of the simplest nontrivial
form with
\begin{gather*}
\lambda^\ms_{i, a} = 0, \qquad (\bm \Psi^\pm_{i, a})^+ = ( \underbracket[.6pt]{1, \ldots, 1}_{i - 1}, (1 - a u)^{\pm 1},
 \underbracket[.6pt]{1, \ldots, 1}_{l - i} ), \quad i \in I, \quad a \in \bbC^\times.
\end{gather*}
Also the one-dimensional representation with the highest $\ell$-weight $\bm \Psi_\xi = (\lambda^\ms_\xi, (\bm \Psi_\xi)^+)$ def\/ined as
\begin{gather}
\lambda_\xi = \xi, \qquad (\bm \Psi_\xi)^+ = \big(q^{\langle \xi, h_1 \rangle}, \ldots q^{\langle \xi, h_l \rangle}\big) \label{psixi}
\end{gather}
is treated as a prefundamental representation. The corresponding $\uqbp$-module is denoted by $L_\xi$. We see that, in the case under consideration, there are actually $2l$ really dif\/ferent prefundamental representations.

It is relevant to recall here the notion of a shifted $\uqbp$-module. Let $V$ be a $\uqbp$-module in the category $\calO$. Given an element $\xi \in \gothh^*$, the shifted $\uqbp$-module $V[\xi]$ is def\/ined as follows. If $\varphi$ is the representation of $\uqbp$ corresponding to the module $V$ and $\varphi[\xi]$ is the representation corresponding to the module $V[\xi]$, then
\begin{gather*}
\varphi[\xi](e_i) = \varphi(e_i), \quad i \in I, \qquad
\varphi[\xi](q^x) = q^{\langle \tilde \xi, x \rangle} \varphi(q^x),
\quad x \in \tgothh.
\end{gather*}
It is clear that the module $V[\xi]$ is in the category $\calO$ and is isomorphic to $V \otimes L_\xi$. Any $\uqbp$-module in the category $\calO$ is a~subquotient of a tensor product of prefundamental representa\-tions~\cite{HerJim12}.

The second example of the highest $\ell$-weight representations of $\uqbp$ is provided by the $q$-oscillator representations. The $q$-oscillator algebra~$\Osc_q$ is a unital associative $\bbC$-algebra with generators $b^\dagger$, $b$, $q^{\nu N}$, $\nu \in \bbC$, satisfying the relations
\begin{gather*}
q^0 = 1, \qquad q^{\nu_1 N} q^{\nu_2 N} = q^{(\nu_1 + \nu_2)N}, \\
q^{\nu N} b^\dagger q^{-\nu N} = q^\nu b^\dagger, \qquad q^{\nu N} b q^{-\nu N} = q^{-\nu} b, \\
b^\dagger b = \frac{q^N - q^{- N}}{q - q^{-1}}, \qquad b b^\dagger = \frac{q q^N - q^{-1} q^{- N}}{q - q^{-1}}.
\end{gather*}
We use two representations of $\Osc_q$. First, let $W^{\scriptscriptstyle +}$ denote the free vector space generated by the set $\{ v_0, v_1, \ldots \}$. The relations
\begin{gather*}
q^{\nu N} v_m = q^{\nu m} v_m, \qquad b^\dagger v_m = v_{m + 1}, \qquad b v_m = [m]_q v_{m - 1},
\end{gather*}
where it is assumed that $v_{-1} = 0$, endow $W^{\scriptscriptstyle +}$ with the structure of an $\Osc_q$-module. The corresponding representation of $\Osc_q$ is denoted by $\chi^{\scriptscriptstyle +}$. Second, let $W^{\scriptscriptstyle -}$ denote the free vector space generated by the set $\{ v_0, v_1, \ldots \}$. The relations
\begin{gather*}
q^{\nu N} v_m = q^{- \nu (m + 1)} v_m, \qquad b v_m = v_{m + 1}, \qquad b^\dagger v_m = - [m]_q v_{m - 1},
\end{gather*}
where it is assumed that $v_{-1} = 0$, endow the vector space $W^{\scriptscriptstyle -}$ with the structure of an $\Osc_q$-module. The corresponding representation of~$\Osc_q$ by~$\chi^{\scriptscriptstyle -}$.

In the case under consideration, we need the tensor product of $l$ copies of the $q$-oscillator algebra, $\Osc_q \otimes \ldots \otimes \Osc_q = (\Osc_q)^{\otimes l}$. Here we introduce the notation{\samepage
\begin{gather*}
\begin{split}
& b_i = 1 \otimes \cdots \otimes 1 \otimes b \otimes 1 \otimes \cdots \otimes 1, \qquad
b_i^\dagger = 1 \otimes \cdots \otimes 1 \otimes b^\dagger \otimes 1 \otimes \cdots \otimes 1, \\
& q^{\nu N_i} = 1 \otimes \cdots \otimes 1 \otimes q^{\nu N} \otimes 1 \otimes \cdots \otimes 1,
\end{split}
\end{gather*}
where $b$, $b^\dagger$ and $q^{\nu N}$ take only the $i$-th place of the respective tensor products.}

As was shown in \cite{NirRaz16b}, the mapping $\rho\colon \uqbp \to (\Osc_q)^{\otimes l}$ def\/ined by the relations
\begin{alignat}{3}
& \rho(q^{\nu h_0}) = q^{\nu \big(2 N_1 + \sum\limits_{j = 2}^l N_j\big)},\qquad && \rho(e_0) = b^\dagger_1 q^{\sum\limits_{j = 2}^l N_j},& \label{roh0e0}\\
& \rho(q^{\nu h_i}) = q^{\nu (N_{i + 1} - N_{i})},\qquad && \rho(e_i) = - b^{}_i b^{\dagger}_{i + 1} q^{N_i - N_{i + 1} - 1},& \label{rohiei}\\
& \rho(q^{\nu h_l}) = q^{- \nu \big(2 N_l + \sum\limits_{j = 1}^{l - 1} N_j\big)}, \qquad && \rho(e_l) = - \kappa_q^{-1} b^\ms_l q^{N_l},&
\label{rohlel}
\end{alignat}
where $i = 1, \ldots, l - 1$, is a homomorphism from the Borel subalgebra~$\uqbp$ to the respective tensor power of the $q$-oscillator algebra. Indeed, relations (\ref{roh0e0})--(\ref{rohlel}) give an obvious interpretation of the $\uqbp$-module relations (\ref{5.4b})--(\ref{5.9b}) in terms of the $q$-oscillators. To get further a representation of~$\uqbp$, one takes the composition of a representation of~$(\Osc_q)^{\otimes l}$ with the mapping $\rho$.

\section{Automorphisms and further representations} \label{s:aafr}

Fixing a f\/inite-dimensional representation of the quantum loop algebra in the quantum space, we can construct explicitly the monodromy- and $L$-operators corresponding to the homomorphisms $\varepsilon \circ \Gamma_\zeta$ and $\rho \circ \Gamma_\zeta$, respectively. Here, $\varepsilon$ is the Jimbo's homomorphism def\/ined in Section~\ref{ss:jh} and~$\rho$ is def\/ined in the preceding section as a homomorphism of~$\uqbp$ to the $q$-oscillator algebra~\cite{NirRaz16b}. Besides, $\Gamma_\zeta$ from Section~\ref{ss:jh} is the grading automorphism of $\uqlsllpo$ introducing the spectral parameter.

Let the f\/inite-dimensional representation in the quantum space be the f\/irst fundamental representation $(\varphi^{(1, 0, \ldots, 0)})_\eta$, so that the monodromy operator is given by the expression
\begin{gather*}
M(\zeta | \eta) = \big(\varepsilon_\zeta \otimes \big(\varphi^{(1, 0, \ldots , 0)}\big)_\eta\big)(\calR).
\end{gather*}
It is clear that $M(\zeta | \eta) \in \uqgllpo \otimes \End(\bbC^{l + 1})$ for any $\zeta, \eta \in \bbC^\times$. It follows from the structure of the universal $R$-matrix that
\begin{gather*}
M(\zeta \nu | \eta \nu) = M(\zeta | \eta).
\end{gather*}
Therefore, one can write
\begin{gather*}
M(\zeta | \eta) = M\big(\zeta \eta^{-1} | 1\big) = M\big(\zeta \eta^{-1}\big),
\end{gather*}
where $M(\zeta) = M(\zeta | 1)$. Identifying $\End(\bbC^{l + 1})$ with $\Mat_{l + 1}(\bbC)$, one can represent
$M(\zeta)$ as
\begin{gather*}
M(\zeta) = \sum_{i, j = 1}^{l + 1} \bbM(\zeta)_{i j} \otimes E_{i j}.
\end{gather*}
Here $\bbM(\zeta)_{i j} \in \uqgllpo$, and $E_{i j} \in \Mat_{l + 1}(\bbC)$ are the standard matrix units.\footnote{Do not confuse with generators of $\uqgllpo$.} We denote by~$\bbM(\zeta)$ the matrix with the matrix entries $\bbM(\zeta)_{i j}$. Generalizing the results of the papers~\cite{NirRaz16a} and~\cite{Raz13}, we see that this matrix has the form
\begin{gather*}
\bbM(\zeta) = q^{K/(l + 1)} \mathrm{e}^{F(\zeta^s)} \tbbM(\zeta),
\end{gather*}
where $F(\zeta)$ is a transcendental function of $\zeta$, while the entries of $\tbbM(\zeta)$ are rational
functions. We use the notation
\begin{gather*}
K = \sum_{i = 1}^{l + 1} K_i, \qquad s = \sum_{i = 0}^l s_i.
\end{gather*}
The function $F(\zeta)$ is def\/ined as follows:
\begin{gather*}
F(\zeta) = \sum_{m \in \mathbb{N}} \frac{C_m}{[l + 1]_{q^m}} \frac{\zeta^m}{m},
\end{gather*}
where one has
\begin{gather*}
\sum_{m \in \mathbb{N}} C_m \frac{\zeta^m}{m} = - \log\left(1 - \sum_{k=1}^{l + 1} C^{(k)} \zeta^k\right),
\end{gather*}
where the elements $C^{(k)}$, $k = 1, \ldots, l + 1$, are the appropriately normalized quantum Casimir operators of the quantum group $\mathrm{U}_q(\mathfrak{gl}_{l + 1})$. For $F(\zeta)$, one also has the def\/ining relations
\begin{gather*}
\sum_{i = 0}^{l} F\big(q^{l - 2 i} \zeta\big) = - \log\left(1 - \sum_{k = 1}^{l + 1} C^{(k)} \zeta^k\right).
\end{gather*}
The of\/f-diagonal matrix entries $\tbbM(\zeta)_{i j}$ are explicitly given by the relations
\begin{alignat*}{3}
& \tbbM(\zeta)_{i j} = - \zeta^{s - s_{ij}} \kappa_q q^{K_i} F_{ij}, \qquad && 1 \le i < j \le l + 1,& \\
& \tbbM(\zeta)_{i j} = - \zeta^{s_{j i}} \kappa_q E_{j i} q^{- K_j}, \qquad && 1 \le j < i \le l + 1,&
\end{alignat*}
while for the diagonal ones we have
\begin{gather*}
\tbbM(\zeta)_{i i} = q^{- K_i} - \zeta^s q^{K_i}, \qquad i = 1, \ldots, l + 1.
\end{gather*}
Here and below we denote
\begin{gather*}
s_{i j} = \sum_{k = i}^{j - 1} s_k.
\end{gather*}

Under the automorphism
\begin{gather*}
E_i \to q^{1/2} E_i q^{K_i - K_{i + 1}}, \qquad F_i \to q^{-1/2} q^{-(K_i - K_{i + 1})} F_i, \qquad q^{\nu K_i} \to q^{\nu K_i}
\end{gather*}
the matrix $\tbbM(\zeta)$ transforms to the matrix for which
\begin{alignat*}{3}
& \tbbM(\zeta)_{i j} = - \zeta^{s - s_{ij}} \kappa_q q^{(K_i + K_j - 1)/2} F_{ij},\qquad && 1 \le i < j \le l + 1,& \\
& \tbbM(\zeta)_{i j} = - \zeta^{s_{j i}} \kappa_q E_{j i} q^{- (K_i + K_j - 1)/2},\qquad && 1 \le j < i \le l + 1.&
\end{alignat*}
The diagonal entries remain the same. Thus, up to a factor belonging to the center of~$\uqgllpo$, we reproduce the result obtained by Jimbo~\cite{Jim86a}.

In a similar way we def\/ine the $L$-operator
\begin{gather*}
L(\zeta | \eta) = \big(\rho_\zeta \otimes\big(\varphi^{(1, 0, \ldots , 0)}\big)_\eta\big)(\calR)
\end{gather*}
and denote by $\bbL(\zeta)$ the corresponding matrix with the entries in $(\Osc_q)^{\otimes l}$. One has
\begin{gather*}
\bbL(\zeta) = \rme^{f(\zeta^s)} \tbbL(\zeta),
\end{gather*}
where the transcendental function $f$ is given by the def\/ining equation
\begin{gather*}
\sum_{j = 0}^{l} f(q^{2j - l} \zeta) = - \log(1 - \zeta)
\end{gather*}
and can explicitly be written as a series
\begin{gather*}
f(\zeta) = \sum_{m \in \bbN} \frac{1}{[l + 1]_{q^m}} \frac{\zeta^m}{m}.
\end{gather*}
The entries of $\tbbL(\zeta)$ are rational functions. For the entries below and above the diagonal we have
\begin{alignat*}{3}
& \tbbL(\zeta)_{i j} = - \zeta^{s_{j i}} \kappa_q b^{}_{j} b^\dagger_{i} q^{N_j + N_{j i} - N_i + i - j - 2}, \qquad && 1 < i - j < l, & \\
& \tbbL(\zeta)_{i + 1, i} = \zeta^{s_{i}} \kappa_q b^{}_{i} b^\dagger_{i + 1} q^{2 N_{i} - N_{i + 1} - 1},\qquad && i = 1,\ldots,l-1,& \\
& \tbbL(\zeta)_{l + 1, i} = \zeta^{s_{i, l + 1}} b^\ms_{i} q^{N_i + N_{i, l + 1} + l - i}, \qquad && i = 1, \ldots, l, &\\
\intertext{and}
& \tbbL(\zeta)_{i, l + 1} = - \zeta^{s - s_{i, l + 1}} \kappa_q b^\dagger_{i} q^{2 N_{1 i} + N_{1, l + 1} + N_{i + 1, l + 1} + i - 1},\qquad && i = 1, \ldots, l, & \\
& \tbbL(\zeta)_{i j} = 0,\qquad && i < j < l + 1, &
\end{alignat*}
respectively.
The diagonal elements of $\tbbL(\zeta)$ are
\begin{gather*}
\tbbL(\zeta)_{i i} = q^{N_i}, \qquad i = 1, \ldots, l, \qquad \tbbL(\zeta)_{l + 1, l + 1} = q^{-N_{1, l + 1}} - \zeta^s q^{N_{1, l + 1} + l + 1}.
\end{gather*}
Here we use the convention
\begin{gather*}
N_{i j} = \sum_{k = i}^{j - 1} N_k.
\end{gather*}
For the cases $l = 1$ and $l = 2$ we refer to the paper \cite{BooGoeKluNirRaz10}, where such $L$-operators were constructed from the universal $R$-matrix.

The monodromy operator $\bbM(\zeta)$ and the $L$-operator $\bbL(\zeta)$ satisfy the Yang--Baxter equation with the $R$-matrix
\begin{gather*}
R(\zeta) = q^{- l / (l + 1)} \rme^{f_{}(q^{-l} \zeta^s) - f_{}(q^{l} \zeta^s)} \widetilde R(\zeta),
\end{gather*}
where
\begin{gather*}
\widetilde R(\zeta) = \sum_{i = 1}^{l+1} E_{ii} \otimes E_{ii} + a(\zeta^s)
\sum_{\substack{i,j = 1 \\ i \neq j}}^{l+1} E_{ii} \otimes E_{jj} \\
\hphantom{\widetilde R(\zeta) =}{} + b(\zeta^s) \left( \sum_{i < j} \zeta^{s_{ij}} E_{ij} \otimes E_{ji}
+ \sum_{i < j} \zeta^{s - s_{ij}} E_{ji} \otimes E_{ij} \right)
\end{gather*}
and we have denoted
\begin{gather*}
a(\zeta) = \frac{q^{}(1 - \zeta)}{1 - q^{2} \zeta}, \qquad b(\zeta) = \frac{1 - q^{2}}{1 - q^{2} \zeta}.
\end{gather*}

Similarly as in Section \ref{ss:jh} more representations of type $\varphi$ were produced by twisting an initial basic representation $(\widetilde \varphi^\lambda)_\zeta$ or $(\varphi^\lambda)_\zeta$ by the automorphisms of~$\uqlsllpo$, also more representations of type $\rho$ can be produced from the initial homomorphism $\rho$ (\ref{roh0e0})--(\ref{rohlel}) twisting it by the automorphisms of~$\uqbp$. The latter can be obtained as the restriction of the automorphisms~$\sigma$ and~$\tau$ from~$\uqlsllpo$ to~$\uqbp$ and are explicitly def\/ined as follows:
\begin{gather*}
\sigma(q^{\nu h_i}) = q^{\nu h_{i + 1}}, \qquad \sigma(e_i) = e_{i + 1}, \qquad i \in \hi,
\end{gather*}
where the identif\/ication $q^{\nu h_{l + 1}} = q^{\nu h_0}$ and $e_{l + 1} = e_0$ is assumed, and
\begin{gather*}
\tau(q^{h_0}) = q^{h_0}, \qquad \tau(q^{h_i}) = q^{h_{l - i + 1}}, \qquad \tau(e_0) = e_0, \qquad \tau(e_i) = e_{l - i + 1}, \qquad i \in I.
\label{taueh}
\end{gather*}
Here we have that $\sigma^{l + 1}$ and $\tau^2$ are the identity transformations. Now we def\/ine
\begin{gather}
\rho_a = \rho \circ \sigma^{-a}, \qquad \ovrho_a = \rho \circ \tau \circ \sigma^{-a+1}, \qquad a = 1, \ldots, l+1,
\label{roaovroa}
\end{gather}
and note that $\ovrho_a$ in (\ref{roaovroa}) can also be written in another form with the help of the relation
\begin{gather*}
\tau \circ \sigma^{-a + 1} = \sigma^{a - 1} \circ \tau = \sigma^{a - l - 2} \circ \tau, \qquad a = 1, \ldots, l+1.
\end{gather*}

Then we obtain from relations (\ref{roh0e0})--(\ref{rohlel})
\begin{alignat}{3}
& \rho_a\big(q^{\nu h_i}\big) = q^{\nu (N_{i - a +1} - N_{i - a})},\qquad && i = a + 1, \ldots, l, \ldots, l + a -1,& \label{rohix} \\
& \rho_a\big(q^{\nu h_{a - 1}}\big) = q^{- \nu \big(2 N_l + \sum\limits_{j=1}^{l-1} N_j\big)},\qquad && \rho_a\big(q^{\nu h_{a}}\big) = q^{\nu \big(2 N_1 + \sum\limits_{j=2}^l N_j\big)},& \label{rohax} \\
& \rho_a(e_i) = - b^{}_{i - a} b^\dagger_{i - a + 1} q^{N_{i - a} - N_{i - a + 1} - 1},\qquad && i = a + 1, \ldots, l, \ldots, l + a -1, & \label{roeix} \\
& \rho_a(e_{a - 1}) = - \kappa_q^{-1} b^{}_{l} q^{N_{l}},\qquad && \rho_a(e_a) = b^{\dagger}_{1} q^{\sum\limits_{j=2}^l N_{j}},& \label{roeax}
\end{alignat}
where $a = 1, \ldots, l+1$, and the index $i$ at the left hand side of (\ref{rohix}) and (\ref{roeix}) takes values modulo $l + 1$. The latter assumption means that the identif\/ication $q^{\nu h_{l+1}} = q^{\nu h_{0}}$ and $e_{l+1} = e_{0}$ holds.

Using tensor products of the representations $\chi^-$ and $\chi^+$, we def\/ine the representations $\theta_a$ as
\begin{gather}
\theta_a = (\underbracket[.6pt]{\chi^- \otimes \cdots \otimes \chi^-}_{l - a + 1} \otimes
\underbracket[.6pt]{\chi^+ \otimes \cdots \otimes \chi^+}_{a - 1}) \circ \rho_a, \qquad a = 1, \ldots, l+1.
\label{thetaa}
\end{gather}
These representations are chosen so as to obtain highest $\ell$-weight representations. The corresponding basis vectors can be def\/ined as
\begin{gather*}
v^{(a)}_{\bm m} = b_1^{m_1} \cdots b_{l - a + 1}^{m_{l-a+1}}
b^{\dagger m_{l-a+2}}_{l - a + 2} \cdots b^{\dagger m_l}_{l}
v_{\bm 0},
\label{vam}
\end{gather*}
where $m_i \in \bbZ_+$ for all $i = 1, \ldots, l$ and we use the notation $\bm m = (m_1, \ldots, m_l)$ and $v_{\bm 0} = v_{(0, \ldots , 0)}$.

For the mappings $\ovrho_a$, $a = 1, \ldots, l+1$, we obtain the following
relations:
\begin{alignat*}{3}
& \ovrho_a\big(q^{\nu h_i}\big) = q^{\nu (N_{a - i} - N_{a - i - 1})}, \qquad && i = 0, 1, \ldots, a - 2,& \\
& \ovrho_a\big(q^{\nu h_{a - 1}}\big) = q^{\nu \big(2 N_1 + \sum\limits_{j=2}^{l} N_j\big)},\qquad
 && \ovrho_a\big(q^{\nu h_{a}}\big) = q^{- \nu \big(2 N_l + \sum\limits_{j=1}^{l-1} N_j\big)}, & \\
& \ovrho_a\big(q^{\nu h_i}\big) = q^{\nu (N_{l + a - i + 1} - N_{l + a - i})}, \qquad && i = a + 1, a + 2, \ldots, l, & \\
& \ovrho_a(e_i) = - b^{}_{a - i - 1} b^\dagger_{a - i} q^{N_{a - i - 1} - N_{a - i} - 1},\qquad && i = 0, 1, \ldots, a - 2,& \\
& \ovrho_a(e_{a - 1}) = b^{\dagger}_{1} q^{\sum\limits_{j=2}^l N_{j}}, \qquad && \ovrho_a(e_a) = - \kappa_q^{-1} b^{}_{l} q^{N_{l}},& \\
& \ovrho_a(e_i) = - b^{}_{l + a - i} b^\dagger_{l + a - i + 1} q^{N_{l + a - i} - N_{l + a - i + 1} - 1},\qquad && i = a + 1, a + 2, \ldots, l.&
\end{alignat*}
Respectively, the homomorphisms $\ovtheta_a$ allowing one to obtain highest $\ell$-weight representations are now def\/ined as
\begin{gather*}
\ovtheta_a = (\underbracket[.6pt]{\chi^- \otimes \cdots \otimes \chi^-}_{a - 1} \otimes \underbracket[.6pt]{\chi^+ \otimes \cdots \otimes \chi^+}_{l - a + 1}) \circ \ovrho_a, \qquad a = 1, \ldots, l+1.\label{ovthetaa}
\end{gather*}
Then the corresponding basis vectors are given by
\begin{gather*}
\ovv^{(a)}_{\bm m} = b_1^{m_1} \cdots b_{a - 1}^{m_{a - 1}} b^{\dagger m_a}_{a \phantom{1}} \cdots b^{\dagger m_l}_l v_{\bm 0}.
\label{ovvam}
\end{gather*}

The vectors $v^{(a)}_{\bm m}$ and $\ovv^{(a)}_{\bm m}$ are actually $\ell$-weight vectors
for the representations $\theta_a$ and $\ovtheta_a$, respectively. Starting from the representations $\theta_a$ and $\ovtheta_a$ we def\/ine the families $(\theta_a)_\zeta$
and $(\ovtheta_a)_\zeta$ as
\begin{gather*}
(\theta_a)_\zeta = \theta_a \circ \Gamma_\zeta, \qquad (\ovtheta_a)_\zeta = \ovtheta_a \circ \Gamma_\zeta.
\end{gather*}
Note here that for a representation $\varphi$ of $\uqlsllpo$ we have
\begin{gather}
\varphi_\zeta(\phi^+_i(u)) = \varphi(\phi^+_i(\zeta^s u)), \qquad \varphi_\zeta\big(\phi^-_i\big(u^{-1}\big)\big) = \varphi\big(\phi^-_i\big(\zeta^{-s} u^{-1}\big)\big). \label{vpiu}
\end{gather}
If $\varphi$ is a representation of $\uqbp$, only the f\/irst one of the above two equations is to be considered. The vectors $v^{(a)}_{\bm m}$ and $\ovv^{(a)}_{\bm m}$ are $\ell$-weight vectors for the representations $(\theta_a)_\zeta$ and $(\ovtheta_a)_\zeta$ as well. We use for the corresponding $\ell$-weights the notation given by the equations
\begin{gather*}
(\theta_a)_\zeta (\phi^+_i(u)) v^{(a)}_{\bm m} = \theta_a(\phi^+_i(\zeta^s u)) v^{(a)}_{\bm m}= \Psi^+_{i, \bm m, a}(u) v^{(a)}_{\bm m}, \\
(\ovtheta_a)_\zeta (\phi^+_i(u)) \ovv^{(a)}_{\bm m} = \ovtheta_a (\phi^+_i(\zeta^s u)) \ovv^{(a)}_{\bm m} = \ovPsi^+_{i, \bm m, a}(u) \ovv^{(a)}_{\bm m},
\end{gather*}
where the f\/irst equation of (\ref{vpiu}) is taken into account. The corresponding elements of $\gothh^*$ are denoted as $\lambda_{\bm m, a}$ and $\ovlambda_{\bm m, a}$. It is worthwhile to note that
\begin{gather*}
\ovrho_a\big(q^{\nu h_i}\big) = \rho_{l - a + 2}\big(q^{\nu h_{l - i + 1}}\big), \qquad \ovrho_a(e_i) = \rho_{l - a + 2}(e_{l - i + 1}) \label{8.b1}
\end{gather*}
and
\begin{gather*}
\bar{v}^{(a)}_{\bm m} = v^{(l - a + 2)}_{\bm m}, \qquad a = 1, \ldots, l + 1.
\end{gather*}
Applied to the relation between $\phi^+_i(u)$ and $e'_{n \delta, \alpha_i}$ in~(\ref{phipiu}), this leads us to the conclusion that the $\ell$-weights $\ovPsi^+_{i, {\bm m}, a}(u)$ are connected with the $\ell$-weights $\Psi^+_{i, {\bm m}, a}(u)$ as
\begin{gather}
\ovPsi^+_{i, {\bm m}, a}(u) = \Psi^+_{l - i + 1, {\bm m}, l - a + 2}\big({-}(-1)^{l} u\big).\label{8.b2}
\end{gather}
For the elements $\ovlambda_{\bm m, a}$ we have
\begin{gather}
\ovlambda_{\bm m, a} = \iota(\lambda_{\bm m, l - a + 2}), \label{blambda}
\end{gather}
where the linear mapping $\iota \colon \gothh^* \to \gothh^*$ is determined by the relation
\begin{gather*}
\iota(\omega_i) = \omega_{l - i + 1}.
\end{gather*}

We have thus $2(l + 1)$ dif\/ferent highest $\ell$-weight $q$-oscillator representations.\footnote{However, for $l = 1$ there are only $2$ such representations.} And this is actually the number of dif\/ferent $L$- and respective $Q$-operators to be considered in the quantum integrable systems associated with the quantum loop algebra $\uqlsllpo$.

\section[Highest $\ell$-weights and functional relations]{Highest $\boldsymbol{\ell}$-weights and functional relations} \label{s:hlwfr}

In our recent paper \cite{BooGoeKluNirRaz17}, we have presented the $\ell$-weights corresponding to the representations $(\theta_a)_\zeta$ and $(\ovtheta_a)_\zeta$, $a = 1, \ldots, l + 1$. As a consequence, putting $\bm m = \bm0$ in those expressions,we obtain the corresponding highest $\ell$-weights.

For the representation $(\theta_a)_\zeta$ we have the highest $\ell$-weights with
\begin{gather}
\lambda_{\bm 0, 1} = -(l + 1) \omega_1, \label{8.1} \\
 \Psi^+_{i, {\bm 0}, 1}(u) =
\begin{cases}
q^{- l - 1} \big( 1 - q^{- l} \zeta^s u \big)^{-1}, & i = 1, \\
1, & i = 2, \ldots, l,
\end{cases} \\
\lambda_{\bm 0, a} = (l - a + 1) \omega_{a - 1} - (l - a + 2) \omega_a, \\
 \Psi^+_{i, {\bm 0}, a}(u) =
\begin{cases}
1, & i = 1,\ldots,a-2, \\
q^{l - a + 1} \big( 1 - q^{- l + a} \zeta^s u \big), & i = a-1, \\
q^{- l + a - 2} \big( 1 - q^{- l + a - 1} \zeta^s u \big)^{-1}, & i = a, \\
1, & i = a+1, \ldots, l,
\end{cases}\label{8.a}
\\
 \lambda_{\bm 0, l + 1} = 0, \\
\Psi^+_{i, {\bm 0}, l + 1}(u) =
\begin{cases}
1, & i = 1, \ldots, l - 1, \\
1 - q \zeta^s u, & i = l.
\end{cases} \label{8.lpo}
\end{gather}
Then, using (\ref{8.b2}) and (\ref{blambda}), we obtain from (\ref{8.1})--(\ref{8.lpo}) the highest $\ell$-weights also for the representations $(\ovtheta_a)_\zeta$. They are
\begin{gather*}
 \ovlambda_{\bm 0, 1} = 0, \\
 \ovPsi^+_{i, {\bm 0}, 1}(u) =
\begin{cases}
1 + (-1)^l q \zeta^s u, & i = 1, \\
1, & i = 2, \ldots, l,
\end{cases}
\\
\ovlambda_{\bm 0, a} = - a \omega_{a - 1} + (a - 1) \omega_a, \\
\ovPsi^+_{i, {\bm 0}, a}(u) =
\begin{cases}
1, & i = 1,\ldots,a-2, \\
q^{- a} \big( 1 + (-1)^l q^{- a + 1} \zeta^s u \big)^{-1}, & i = a-1, \\
q^{a - 1} \big( 1 + (-1)^l q^{- a + 2} \zeta^s u \big), & i = a, \\
1, & i = a+1, \ldots, l,
\end{cases}
\\
 \ovlambda_{\bm 0, l + 1} = {} - (l + 1) \omega_l, \\
 \ovPsi^+_{i, {\bm 0}, l + 1}(u) =
\begin{cases}
1, & i = 1, \ldots, l - 1, \\
q^{- l - 1}\big( 1 + (-1)^l q^{- l} \zeta^s u \big)^{-1}, & i = l.
\end{cases}
\end{gather*}

\looseness=1 The explicit forms of the highest $\ell$-weights allow us to conclude that the representa\-tions $(\theta_{l+1})_\zeta$ and~$(\ovtheta_1)_\zeta$ are isomorphic to prefundamental representations, the representa\-tions~$(\theta_1)_\zeta$ and $(\ovtheta_{l+1})_\zeta$ are isomorphic to shifted prefundamental representations, and the other representa\-tions~$(\theta_a)_\zeta$ and~$(\ovtheta_a)_\zeta$ with $a = 2, \ldots, l$ are isomorphic to subquotients of tensor products of two certain prefundamental representations of~$\uqbp$. Explicitly we have
\begin{gather*}
(\theta_1)_\zeta \cong L^\ms_{\xi_1} \otimes L^-_{1, q^{- l} \zeta^s} , \\
(\theta_a)_\zeta \cong L^\ms_{\xi_a} \otimes \big(L^+_{a - 1, q^{- l + a} \zeta^s} \ovotimes L^-_{a, q^{- l + a - 1} \zeta^s }\big), \qquad a = 2, \ldots, l, \\
(\theta_{l + 1})_\zeta \cong L^+_{l, q \zeta^s},
\end{gather*}
where the shifts $\xi_{a}$ are determined by the equation
\begin{gather*}
\xi_a = (l - a + 1) \omega_{a - 1} - (l - a + 2) \omega_a
\end{gather*}
for the representations $(\theta_a)_\zeta$, and
\begin{gather*}
\big(\ovtheta_1\big)_\zeta \cong L^+_{1, (-1)^{l + 1} q^{} \zeta^s} , \\
\big(\ovtheta_a\big)_\zeta \cong L^\ms_{\bar\xi_a} \otimes \big(L^-_{a - 1, (-1)^{l - 1} q^{- a + 1} \zeta^s} \ovotimes L^+_{a, (-1)^{l + 1} q^{- a + 1} \zeta^s }\big), \qquad a = 2, \ldots, l, \\
\big(\ovtheta_{l + 1}\big)_\zeta \cong L^\ms_{\bar\xi_{l+1}} \otimes L^-_{l, (-1)^{l + 1} q^{- l} \zeta^s},
\end{gather*}
where the shifts $\bar\xi_a$ are determined by the equation
\begin{gather*}
\ovxi_a = {} - a \omega_{a - 1} + (a - 1) \omega_a
\end{gather*}
for the representations $(\ovtheta_a)_\zeta$. The operation $\ovotimes$ means taking the corresponding subquotients, as introduced in Section~\ref{s:hlwr}.

We can also reverse the above relations in order to express the prefundamental representations via subquotients of tensor products of highest $\ell$-weight $q$-oscillator representations. We obtain
\begin{gather}
L_{\xi^-_i} \otimes L^-_{i, \zeta^s} \cong (\theta_1)_{q^{l + i - 1} \zeta^s}
\ovotimes (\theta_2)_{q^{l + i - 3} \zeta^s} \ovotimes \cdots \ovotimes (\theta_i)_{q^{l - i + 1} \zeta^s},
\label{L-1}
\end{gather}
where the elements $\xi^-_i$ are def\/ined as
\begin{gather}
\xi^-_i = - 2 \sum_{j = 1}^{i - 1} \omega_j - (l - i + 2) \omega_i, \label{xi-1}
\end{gather}
and
\begin{gather}
L_{\xi^+_i} \otimes L^+_{i, \zeta^s} \cong (\theta_{i + 1})_{q^{l - i - 1} \zeta^s}
\ovotimes (\theta_{i + 2})_{q^{l - i - 3} \zeta^s} \ovotimes \cdots \ovotimes (\theta_{l + 1})_{q^{- l + i - 1} \zeta^s},
\label{L+1}
\end{gather}
with the elements $\xi^+_i$ given by the equation
\begin{gather}
\xi^+_i = (l - i) \omega_i - 2 \sum_{j = i + 1}^j \omega_j.
\label{xi+1}
\end{gather}

Similar relations can also be written for the representations $\ovtheta_a$. Actually, we have
\begin{gather}
L_{\xi^+_i} \otimes L^+_{i, \zeta^s} \cong \big(\ovtheta_1\big)_{(-1)^{l-1} q^{- i} \zeta^s} \ovotimes \big(\ovtheta_2\big)_{(-1)^{l-1} q^{2 - i} \zeta^s} \ovotimes \cdots \ovotimes \big(\ovtheta_i\big)_{(-1)^{l-1} q^{i - 2} \zeta^s},\label{L+2}
\end{gather}
where the elements $\xi^+_i$ are def\/ined as
\begin{gather}
\xi^+_i = - 2 \sum_{j = 1}^{i - 1} \omega_j + (i - 1) \omega_i \label{xi+2}
\end{gather}
and
\begin{gather}
L_{\xi^-_i} \otimes L^-_{i, \zeta^s} \cong \big(\ovtheta_{i + 1}\big)_{(-1)^{l-1} q^{i} \zeta^s}
\ovotimes \big(\ovtheta_{i + 2}\big)_{(-1)^{l-1} q^{i + 2} \zeta^s} \ovotimes \cdots \ovotimes
\big(\ovtheta_{l + 1}\big)_{(-1)^{l-1} q^{2 l - i} \zeta^s},\label{L-2}
\end{gather}
with the elements $\xi^-_i$ given by the equation
\begin{gather}
\xi^-_i = - (i + 1) \omega_i - 2 \sum_{j = i + 1}^l \omega_j.\label{xi-2}
\end{gather}

Now it is clear that, in the case under consideration, the $q$-oscillator representations could quite be treated as no less fundamental than the prefundamental ones. Indeed, any $\uqbp$-module in the category~$\calO$ can be presented as a shifted subquotient of a~tensor product of $q$-oscillator representations. And it should also be noted that the highest $\ell$-weights of the $q$-oscillator representations are as simple as
the highest $\ell$-weights of the prefundamental representations.

We denote the $\uqbp$-modules corresponding to the representations $\theta_a$ def\/ined in~(\ref{thetaa}) by~$W_a$, $a = 1, \ldots, l+1$, and consider the $\uqbp$-module $(W_1)_{\zeta_1} \otimes \cdots \otimes (W_{l+1})_{\zeta_{l+1}}$. Then, the tensor product of the highest $\ell$-weight vectors is an $\ell$-weight vector of $\ell$-weight determined by the functions
\begin{gather*}
\Psi^+_i(u) = q^{-2} \frac{1 - q^{- l + i + 1} \zeta^s_{i+1} u} {1 - q^{- l + i - 1} \zeta^s_{i} u}, \qquad i = 1, \ldots, l.
\end{gather*}
And now, let us take the representation $\widetilde \varphi^\lambda$, $\lambda \in \gothk^*$, of the whole quantum loop algebra $\uqlsllpo$ constructed with the help of the Jimbo's homomorphism \cite{NirRaz16b} and consider its restriction to the Borel subalgebra~$\uqbp$. We denote this restriction and the corresponding $\uqbp$-module again by $\widetilde \varphi^\lambda$ and $\widetilde V^\lambda$. It can be shown\footnote{Work in progress, to appear elsewhere.} that the highest $\ell$-weight of the $\uqbp$-module $(\widetilde V^\lambda)_\zeta$ is determined by the functions
\begin{gather*}
\Psi^+_i(u) = q^{\lambda_i - \lambda_{i+1}} \frac{1 - q^{2\lambda_{i+1} - i + 1} \zeta^s u} {1 - q^{2\lambda_{i} - i + 1} \zeta^s u}, \qquad
i = 1, \ldots, l.
\end{gather*}
Let $\rho$ denote the half-sum of all positive roots of $\gothgl_{l + 1}$. One can show that
\begin{gather*}
\langle \rho, K_i \rangle = \frac{l}{2} - i + 1.
\end{gather*}
We see that if
\begin{gather*}
\zeta_i = q^{2 \langle \lambda + \rho, K_i \rangle / s} \zeta,
\end{gather*}
then the submodule of $(W_1)_{\zeta_1} \otimes \cdots \otimes (W_{l+1})_{\zeta_{l+1}}$ generated by the tensor product of the highest $\ell$-weight vectors of $(W_a)_{\zeta_a}$, $a = 1, \ldots, l + 1$, is isomorphic to the shifted module $(\widetilde V^\lambda)_\zeta[\xi]$, where the shift $\xi$ is determined by the equation
\begin{gather*}
\xi = - \sum_{i = 1}^l (\lambda_i - \lambda_{i + 1} + 2) \omega_i.
\end{gather*}
A similar conclusion holds for $\ovtheta_a$ as well. This connection between the highest $\ell$-weights ref\/lects the basic functional relation between the universal transfer operator based on the inf\/inite-dimensional representation $(\widetilde\varphi^\lambda)_\zeta$ and the product of all universal $Q$-operators based on the $q$-oscillator representations $(\theta_a)_\zeta$ at certain values of the spectral parameters. Such relations for $l = 1$ and $l = 2$ were proved in the papers \cite{BazLukZam97, BazLukZam99, BooGoeKluNirRaz13, BooGoeKluNirRaz14a, NirRaz16a} and \cite{BazHibKho02, BooGoeKluNirRaz14b}, respectively.

Besides, comparing (\ref{L-1}) with (\ref{L-2}), also taking into account the shifts (\ref{xi-1}) and (\ref{xi-2}), we can relate the integrability objects $\calQ_a(\zeta)$ and $\ovcalQ_a(\zeta)$ associated with the representations $(\theta_a)_\zeta$ and $(\ovtheta_a)_\zeta$, respectively. Specif\/ically, linear combinations of the products $\calQ_1(\zeta_1) \cdots \calQ_i(\zeta_i)$ are expressed through the product $\ovcalQ_{i+1}(\zeta_{i+1}) \cdots \ovcalQ_{l+1}(\zeta_{l+1})$, $i = 1, \ldots, l$, at certain values of the spectral parameters $\zeta_a$, $a = 1, \ldots, l + 1$. In the same way, comparing (\ref{L+1}) with (\ref{L+2}), also taking into account the shifts (\ref{xi+1}) and (\ref{xi+2}), we can relate the products $\ovcalQ_1(\zeta_1) \cdots \ovcalQ_i(\zeta_i)$ with the products $\calQ_{i+1}(\zeta_{i+1}) \cdots \calQ_{l+1}(\zeta_{l+1})$, $i = 1, \ldots, l$, at certain values of the spectral parameters $\zeta_a$, $a = 1, \ldots, l + 1$. Obviously, such relations between the universal $Q$-operators are absent if $l = 1$. For the simplest higher rank case, $l = 2$, the corresponding relations were proved in \cite{BazHibKho02, BooGoeKluNirRaz14b}.

\section{Conclusion}

We have explicitly related the highest $\ell$-weight $q$-oscillator representations $(\theta_a)_\zeta$, $(\ovtheta_a)_\zeta$ of the Borel subalgebra $\uqbp$ of the quantum loop algebra $\uqlsllpo$ with arbitrary rank $l$ with the shifted prefundamental representations $L^\pm_{i, x}$ from the category~$\calO$. Thus, not only the $q$-oscillator representations can be obtained as subquotients of tensor products of the prefundamental representations, but also the latter can be expressed via appropriate tensor products of the former. We have also demonstrated how the information about $\ell$-weights can be used for the construction of functional relations.

For all representations $(\theta_a)_\zeta$, $(\ovtheta_a)_\zeta$ there is a~basis of the corresponding representation space consisting of $\ell$-weight vectors~\cite{BooGoeKluNirRaz17}. It is worthwhile noting that for $l \ge 2$ one has $2(l + 1)$ $q$-oscillator representations and only~$2l$ prefundamental representations. The case of $l = 1$ is special in the sense that only~$2$ representations of both kinds are present~\cite{BooGoeKluNirRaz16}.

\subsection*{Acknowledgements} We are grateful to H.~Boos, F.~G\"ohmann and A.~Kl\"umper for discussions. This work was supported in part by the Deutsche Forschungsgemeinschaft in the framework of the research group FOR 2316, by the DFG grant KL \hbox{645/10-1}, and by the RFBR grants \#~14-01-91335 and \#~16-01-00473. Kh.S.N.\ is grateful to the RAQIS'16 Organizers for the invitation and hospitality during the Conference ``Recent Advances in Quantum Integrable Systems'', August 22--26, 2016, at the University of Geneva.

\pdfbookmark[1]{References}{ref}
\LastPageEnding


\begin{thebibliography}{99}
\footnotesize\itemsep=0pt

\bibitem{AshSmiTol79}
Asherova R.M., Smirnov Yu.F., Tolstoy V.N., Description of a class of projection
 operators for semisimple complex {L}ie algebras, \href{https://doi.org/10.1007/BF01140268}{\textit{Math. Notes}}
 \textbf{26} (1979), 499--504.

\bibitem{BazHibKho02}
Bazhanov V.V., Hibberd A.N., Khoroshkin S.M., Integrable structure of
 {${\mathcal W}_3$} conformal f\/ield theory, quantum {B}oussinesq theory and
 boundary af\/f\/ine {T}oda theory, \href{https://doi.org/10.1016/S0550-3213(01)00595-8}{\textit{Nuclear Phys.~B}} \textbf{622} (2002),
 475--547, \href{https://arxiv.org/abs/hep-th/0105177}{hep-th/0105177}.

\bibitem{BazLukZam96}
Bazhanov V.V., Lukyanov S.L., Zamolodchikov A.B., Integrable structure of
 conformal f\/ield theory, quantum {K}d{V} theory and thermodynamic {B}ethe
 ansatz, \href{https://doi.org/10.1007/BF02101898}{\textit{Comm. Math. Phys.}} \textbf{177} (1996), 381--398,
 \href{https://arxiv.org/abs/hep-th/9412229}{hep-th/9412229}.

\bibitem{BazLukZam97}
Bazhanov V.V., Lukyanov S.L., Zamolodchikov A.B., Integrable structure of
 conformal f\/ield theory. {II}.~{${\rm Q}$}-ope\-rator and {DDV} equation,
 \href{https://doi.org/10.1007/s002200050240}{\textit{Comm. Math. Phys.}} \textbf{190} (1997), 247--278,
 \href{https://arxiv.org/abs/hep-th/9604044}{hep-th/9604044}.

\bibitem{BazLukZam99}
Bazhanov V.V., Lukyanov S.L., Zamolodchikov A.B., Integrable structure of
 conformal f\/ield theory. {III}.~{T}he {Y}ang--{B}axter relation, \href{https://doi.org/10.1007/s002200050531}{\textit{Comm.
 Math. Phys.}} \textbf{200} (1999), 297--324, \href{https://arxiv.org/abs/hep-th/9805008}{hep-th/9805008}.

\bibitem{Beck94}
Beck J., Braid group action and quantum af\/f\/ine algebras, \href{https://doi.org/10.1007/BF02099423}{\textit{Comm. Math.
 Phys.}} \textbf{165} (1994), 555--568, \mbox{\href{https://arxiv.org/abs/hep-th/9404165}{hep-th/9404165}}.

\bibitem{BooGoeKluNirRaz10}
Boos H., G\"ohmann F., Kl\"umper A., Nirov Kh.S., Razumov A.V., Exercises with
 the universal {$R$}-matrix, \href{https://doi.org/10.1088/1751-8113/43/41/415208}{\textit{J.~Phys.~A: Math. Theor.}} \textbf{43}
 (2010), 415208, 35~pages, \href{https://arxiv.org/abs/1004.5342}{arXiv:1004.5342}.

\bibitem{BooGoeKluNirRaz13}
Boos H., G\"ohmann F., Kl\"umper A., Nirov Kh.S., Razumov A.V., Universal
 integrability objects, \href{https://doi.org/10.1007/s11232-013-0002-8}{\textit{Theoret. and Math. Phys.}} \textbf{174} (2013),
 21--39, \href{https://arxiv.org/abs/1205.4399}{arXiv:1205.4399}.

\bibitem{BooGoeKluNirRaz14b}
Boos H., G\"ohmann F., Kl\"umper A., Nirov Kh.S., Razumov A.V., Quantum groups
 and functional relations for higher rank, \href{https://doi.org/10.1088/1751-8113/47/27/275201}{\textit{J.~Phys.~A: Math. Theor.}}
 \textbf{47} (2014), 275201, 47~pages, \href{https://arxiv.org/abs/1312.2484}{arXiv:1312.2484}.

\bibitem{BooGoeKluNirRaz14a}
Boos H., G\"ohmann F., Kl\"umper A., Nirov Kh.S., Razumov A.V., Universal
 {$R$}-matrix and functional relations, \href{https://doi.org/10.1142/S0129055X14300052}{\textit{Rev. Math. Phys.}} \textbf{26}
 (2014), 1430005, 66~pages, \href{https://arxiv.org/abs/1205.1631}{arXiv:1205.1631}.

\bibitem{BooGoeKluNirRaz16}
Boos H., G\"ohmann F., Kl\"umper A., Nirov Kh.S., Razumov A.V., Oscillator
 versus prefundamental representations, \href{https://doi.org/10.1063/1.4966925}{\textit{J.~Math. Phys.}} \textbf{57}
 (2016), 111702, 23~pages, \href{https://arxiv.org/abs/1512.04446}{arXiv:1512.04446}.

\bibitem{BooGoeKluNirRaz17}
Boos H., G\"ohmann F., Kl\"umper A., Nirov Kh.S., Razumov A.V., Oscillator
 versus prefundamental representations. {II}.~{A}rbitrary higher ranks,
 \href{https://arxiv.org/abs/1701.0262}{arXiv:1701.0262}.

\bibitem{ChaPre91}
Chari V., Pressley A., Quantum af\/f\/ine algebras, \href{https://doi.org/10.1007/BF02102063}{\textit{Comm. Math. Phys.}}
 \textbf{142} (1991), 261--283.

\bibitem{ChaPre94}
Chari V., Pressley A., A guide to quantum groups, Cambridge University Press,
 Cambridge, 1994.

\bibitem{Dam12}
Damiani I., Drinfeld realization of af\/f\/ine quantum algebras: the relations,
 \href{https://doi.org/10.2977/PRIMS/86}{\textit{Publ. Res. Inst. Math. Sci.}} \textbf{48} (2012), 661--733,
 \href{https://arxiv.org/abs/1406.6729}{arXiv:1406.6729}.

\bibitem{Dam15}
Damiani I., From the {D}rinfeld realization to the {D}rinfeld--{J}imbo
 presentation of af\/f\/ine quantum algebras: injectivity, \href{https://doi.org/10.4171/PRIMS/150}{\textit{Publ. Res.
 Inst. Math. Sci.}} \textbf{51} (2015), 131--171, \href{https://arxiv.org/abs/1407.0341}{arXiv:1407.0341}.

\bibitem{Dri85}
Drinfel'd V.G., Hopf algebras and the quantum {Y}ang--{B}axter equation,
 \textit{Sov. Math. Dokl.} \textbf{32} (1985), 254--258.

\bibitem{Dri87}
Drinfel'd V.G., Quantum groups, in Proceedings of the {I}nternational
 {C}ongress of {M}athematicians, {V}ols.~1,~2 ({B}erkeley, {C}alif., 1986),
 Amer. Math. Soc., Providence, RI, 1987, 798--820.

\bibitem{Dri88}
Drinfel'd V.G., A new realization of {Y}angians and of quantum af\/f\/ine algebras,
 \textit{Sov. Math. Dokl.} \textbf{36} (1988), 212--216.

\bibitem{EtiFreKir98}
Etingof P.I., Frenkel I.B., Kirillov Jr. A.A., Lectures on representation
 theory and {K}nizhnik--{Z}amolodchikov equations, \href{https://doi.org/10.1090/surv/058}{\textit{Mathematical
 Surveys and Monographs}}, Vol.~58, Amer. Math. Soc., Providence, RI, 1998.

\bibitem{FeiJimMiwMuk16}
Feigin B., Jimbo M., Miwa T., Mukhin E., Finite type modules and {B}ethe ansatz
 equations, \href{https://arxiv.org/abs/1609.05724}{arXiv:1609.05724}.

\bibitem{FreHer15}
Frenkel E., Hernandez D., Baxter's relations and spectra of quantum integrable
 models, \href{https://doi.org/10.1215/00127094-3146282}{\textit{Duke Math.~J.}} \textbf{164} (2015), 2407--2460,
 \href{https://arxiv.org/abs/1308.3444}{arXiv:1308.3444}.

\bibitem{FreRes99}
Frenkel E., Reshetikhin N., The {$q$}-characters of representations of quantum
 af\/f\/ine algebras and deformations of {$\mathcal W$}-algebras, in Recent
 Developments in Quantum Af\/f\/ine Algebras and Related Topics ({R}aleigh, {NC},
 1998), \href{https://doi.org/10.1090/conm/248/03823}{\textit{Contemp. Math.}}, Vol.~248, Amer. Math. Soc., Providence, RI,
 1999, 163--205, \href{https://arxiv.org/abs/math.QA/9810055}{math.QA/9810055}.

\bibitem{Her04}
Hernandez D., Representations of quantum af\/f\/inizations and fusion product,
 \href{https://doi.org/10.1007/s00031-005-1005-9}{\textit{Transform. Groups}} \textbf{10} (2005), 163--200,
 \href{https://arxiv.org/abs/math.QA/0312336}{math.QA/0312336}.

\bibitem{Her07}
Hernandez D., Drinfeld coproduct, quantum fusion tensor category and
 applications, \href{https://doi.org/10.1112/plms/pdm017}{\textit{Proc. Lond. Math. Soc.}} \textbf{95} (2007), 567--608,
 \href{https://arxiv.org/abs/math.QA/0504269}{math.QA/0504269}.

\bibitem{HerJim12}
Hernandez D., Jimbo M., Asymptotic representations and {D}rinfeld rational
 fractions, \href{https://doi.org/10.1112/S0010437X12000267}{\textit{Compos. Math.}} \textbf{148} (2012), 1593--1623,
 \href{https://arxiv.org/abs/1104.1891}{arXiv:1104.1891}.

\bibitem{Jim85}
Jimbo M., A {$q$}-dif\/ference analogue of {$U({\mathfrak g})$} and the
 {Y}ang--{B}axter equation, \href{https://doi.org/10.1007/BF00704588}{\textit{Lett. Math. Phys.}} \textbf{10} (1985),
 63--69.

\bibitem{Jim86a}
Jimbo M., A {$q$}-analogue of {$U({\mathfrak g}{\mathfrak l}(N+1))$}, {H}ecke
 algebra, and the {Y}ang--{B}axter equation, \href{https://doi.org/10.1007/BF00400222}{\textit{Lett. Math. Phys.}}
 \textbf{11} (1986), 247--252.

\bibitem{JimMiw95}
Jimbo M., Miwa T., Algebraic analysis of solvable lattice models, \href{https://doi.org/10.1090/cbms/085}{\textit{CBMS
 Regional Conference Series in Mathematics}}, Vol.~85, Amer. Math. Soc.,
 Providence, RI, 1995.

\bibitem{Kac90}
Kac V.G., Inf\/inite-dimensional {L}ie algebras, 3rd ed., \href{https://doi.org/10.1017/CBO9780511626234}{Cambridge University
 Press}, Cambridge, 1990.

\bibitem{KhoTol93}
Khoroshkin S.M., Tolstoy V.N., On {D}rinfel'd's realization of quantum af\/f\/ine
 algebras, \href{https://doi.org/10.1016/0393-0440(93)90070-U}{\textit{J.~Geom. Phys.}} \textbf{11} (1993), 445--452.

\bibitem{KhoTol94}
Khoroshkin S.M., Tolstoy V.N., Twisting of quantum (super)algebras.
 {C}onnection of {D}rinfeld's and {C}artan--{W}eyl realizations for quantum
 af\/f\/ine algebras, \href{https://arxiv.org/abs/hep-th/9404036}{hep-th/9404036}.

\bibitem{LezSav74}
Leznov A.N., Savel'ev M.V., A parametrization of compact groups, \href{https://doi.org/10.1007/BF01075497}{\textit{Funct.
 Anal. Appl.}} \textbf{8} (1974), 347--348.

\bibitem{MenTes15}
Meneghelli C., Teschner J., Integrable light-cone lattice discretizations from
 the universal ${R}$-matrix, \href{https://arxiv.org/abs/1504.04572}{arXiv:1504.04572}.

\bibitem{MukYou14}
Mukhin E., Young C.A.S., Af\/f\/inization of category {$\mathcal{O}$} for quantum
 groups, \href{https://doi.org/10.1090/S0002-9947-2014-06039-X}{\textit{Trans. Amer. Math. Soc.}} \textbf{366} (2014), 4815--4847,
 \href{https://arxiv.org/abs/1204.2769}{arXiv:1204.2769}.


\bibitem{NirRaz16a}
Nirov Kh.S., Razumov A.V., Quantum groups and functional relations for lower
 rank, \href{https://doi.org/10.1016/j.geomphys.2016.10.014}{\textit{J.~Geom. Phys.}} \textbf{112} (2017), 1--28, \href{https://arxiv.org/abs/1412.7342}{arXiv:1412.7342}.

\bibitem{NirRaz16b}
Nirov Kh.S., Razumov A.V., Quantum groups, {V}erma modules and $q$-oscillators:
 General linear case, \href{https://arxiv.org/abs/1610.02901}{arXiv:1610.02901}.

\bibitem{Raz13}
Razumov A.V., Monodromy operators for higher rank, \href{https://doi.org/10.1088/1751-8113/46/38/385201}{\textit{J.~Phys.~A: Math.
 Theor.}} \textbf{46} (2013), 385201, 24~pages, \href{https://arxiv.org/abs/1211.3590}{arXiv:1211.3590}.

\bibitem{TolKho92}
Tolstoy V.N., Khoroshkin S.M., The universal {$R$}-matrix for quantum untwisted
 af\/f\/ine {L}ie algebras, \href{https://doi.org/10.1007/BF01077085}{\textit{Funct. Anal. Appl.}} \textbf{26} (1992),
 69--71.

\bibitem{Yam89}
Yamane H., A {P}oincar\'e--{B}irkhof\/f--{W}itt theorem for quantized universal
 enveloping algebras of type~{$A_N$}, \href{https://doi.org/10.2977/prims/1195173355}{\textit{Publ. Res. Inst. Math. Sci.}}
 \textbf{25} (1989), 503--520.

\end{thebibliography}
\end{document}